\documentclass{aa}

\usepackage{graphicx}
\usepackage[svgnames]{xcolor}
\usepackage[pdftex,colorlinks,citecolor=DarkBlue]{hyperref}

\usepackage[varg]{txfonts}

\usepackage{natbib}
\bibpunct{(}{)}{;}{a}{}{,} % to follow the A&A style

\begin{document}

   \title{ Structure distribution and turbulence in  self-consistently supernova-driven
   ISM of multiphase magnetized galactic discs}

   \titlerunning{Turbulence and structure formation in the ISM}

   \author{Olivier Iffrig\inst{\ref{inst1}}
      \and
      Patrick Hennebelle\inst{\ref{inst1},\ref{inst2}}}

   \institute{Laboratoire AIM, Paris-Saclay, CEA/IRFU/SAp -- CNRS --
      Universit\'e Paris Diderot, 91191 Gif-sur-Yvette Cedex, France
      \label{inst1}
      \and
      LERMA (UMR CNRS 8112), Ecole Normale Sup\'erieure, 75231 Paris Cedex,
      France
      \label{inst2}
      }

   \date{Received ---; accepted ---}

% \abstract{}{}{}{}{}
% 5 {} token are mandatory

   \abstract%
      % context heading (optional), leave it empty if necessary
      {Galaxy evolution and star formation are two multi-scale problems tightly linked to each other.}
      {To understand the interstellar cycle, which  triggers galaxy evolution, it is necessary 
      to describe simultaneously the large-scale evolution widely induced by the feedback processes and the 
      details of the gas dynamics that controls the star formation process through gravitational collapse. }
      {%
         We perform a set of three-dimensional high-resolution numerical simulations of a
         turbulent, self-gravitating and magnetized interstellar medium within a
         $1\ \mathrm{kpc}$ stratified box with supernova feedback correlated with
         star-forming regions. In particular, we focus on the role played by the 
         magnetic field and the feedback on the galactic vertical structure, the star formation rate (SFR)
         and the flow dynamics. For this purpose we vary their respective intensities.
         We extract properties of the dense clouds arising
         from the turbulent motions and compute power spectra of various quantities.
         %We compute both three- and two-dimensional power spectra.%
      }
      {%
       Using a distribution of supernovae sufficiently correlated with the dense gas, we  
       find that supernova explosions can reproduce the observed SFR, particularly if the 
       magnetic field is on the order of a few $\mu G$. The vertical structure, which results
       from a dynamical and an energy equilibrium is well reproduced by a simple analytical model, 
       which allows us to estimate the coupling between the gas and the supernovae. We  found 
       the coupling to be rather low and on the order of 1.5$\%$. Strong magnetic fields may help   
       to increase this coupling by a factor of about 2-3.
       To characterize the flow we compute the power spectra of various quantities in 3D
       but also in 2D in order to account for the stratification of the galactic disc. 
       Noticeably we find that within our setup, the compressive modes tend to dominate 
       in the equatorial plane, while at about one scale height above it, solenoidal modes
       become dominant. We measure the angle between the magnetic and  velocity fields 
       and we conclude that they tend to be well aligned particularly at high magnetization
       and lower feedback. Finally, the dense structures present scaling relations 
       that are reminiscent of the observational ones. The  
        virial parameter is typically larger than 10 and shows a large spread of  masses below 1000 M$_\odot$.
      For masses larger than 10$^4$ M$_\odot$, its value tends to a few. 
      }
      {Using a relatively simple scheme for the supernova feedback, which is 
      self-consistently proportional  to the SFR and spatially correlated to the 
      star formation process, we reproduce a stratified galactic disc that
      presents reasonable scale height, SFR as well as a cloud distribution
      with characteristics close to the observed ones.}


   \maketitle

%________________________________________________________________

\section{Introduction}

Understanding the cycle of matter and energy within galaxies is a necessary 
step for our knowledge of the structure formation in the  universe. 
Because of the wide diversity of spatial and temporal scales, which govern
these cycles, it is not possible to directly simulate a galaxy 
\citep[e.g.][]{tasker-bryan2006,dubois+2008,bournaud+2010,kim+2011,dobbs+2011,tasker2011,hopkins+2011,renaud+2013}, 
with a 
well-resolved interstellar medium (ISM) although the spatial resolution is 
continuously improving.
To address this question, an alternative approach has been developed which consists in 
simulating a small portion of a galactic disc leading to a better spatial resolution
\citep{korpi+1999,slyz+2005,deavillez+2005,Joung06,Hill12, kim+2011, kim+2013, gent+2013, Hennebelle14,gatto2015}
although at the expense of not solving the large galactic scales. Clearly these two approaches 
are complementary and must be used in parallel.

Generally speaking the most important motivation for performing this type of calculations is to study 
the impact of various feedback mechanisms such as  supernovae on  
the star formation rate (SFR), the structure of the galactic disc and the outflows that are launched. 
Due to technical difficulties, in particular the very small time steps induced by the 
high velocities and temperatures that feedback generates, these studies are usually limited 
to relatively modest numerical spatial resolution and as a consequence the detailed flow properties have been 
more rarely considered \citep{Deavillez2007,Padoan16}, being mainly addressed through smaller
scale simulations in which the energy input has to be prescribed \citep{vazquez2006,Kritsuk07,Banerjee09,Audit2010,Inoue2012}. 

In this paper we continue on our previous 
study \citep{Hennebelle14} by performing a series of high-resolution simulations varying the 
magnetic intensities from purely hydrodynamical to strongly magnetized flows. In order to 
better constrain the influence that feedback exerts onto  the galactic disc evolution, we also 
reduce its efficiency and compare with the fiducial feedback values.
Our aim is to characterize not only the  SFR and disc structure but also the properties of the turbulence which 
develops. Indeed, the flows in such an  environment are complex because not only they are self-gravitating 
and magnetized but also because energy is injected at intermediate scales through the stellar feedback processes, 
and last but not least, because the galactic discs are strongly stratified making them highly non isotropic. 
Characterizing the turbulence and more generally the physical and  statistical properties of these flows is a necessary step
to understand the star formation process and more generally the energy cycles that are taking place in the 
galaxies. Therefore we calculate the various bidimensional power spectra at various altitudes finding for some of them
a strong dependence with it. We also investigate the alignment between the velocity and the magnetic field. 
Indeed, not only this latter may be playing a role in the theory of MHD turbulence \citep{Boldyrev06} but also
the strong influence it may have on the molecular cloud evolution and collapse has been recently 
stressed by \citet{Inoue09} and \citet{kortgen2015}.
 Finally, we  extract the clumps and study their properties in particular how they depend 
on the magnetization and the feedback strength.

In the second section of the paper we describe the numerical setup that we use. In the
third section we give a general description and analysis of the mean quantities such as the star formation rate 
as well as the thickness of the galactic plane. In the fourth section, we study the cold clumps
extracted from the simulations and discuss their physical properties. The fifth section is dedicated to 
the physical properties of the turbulence which develops in our simulations and we study the 
power spectra, the alignment between the velocity and magnetic field as well as the structure properties. The sixth section 
concludes the paper.

      \begin{figure*}
         \begin{center}
            \includegraphics[width=15cm]{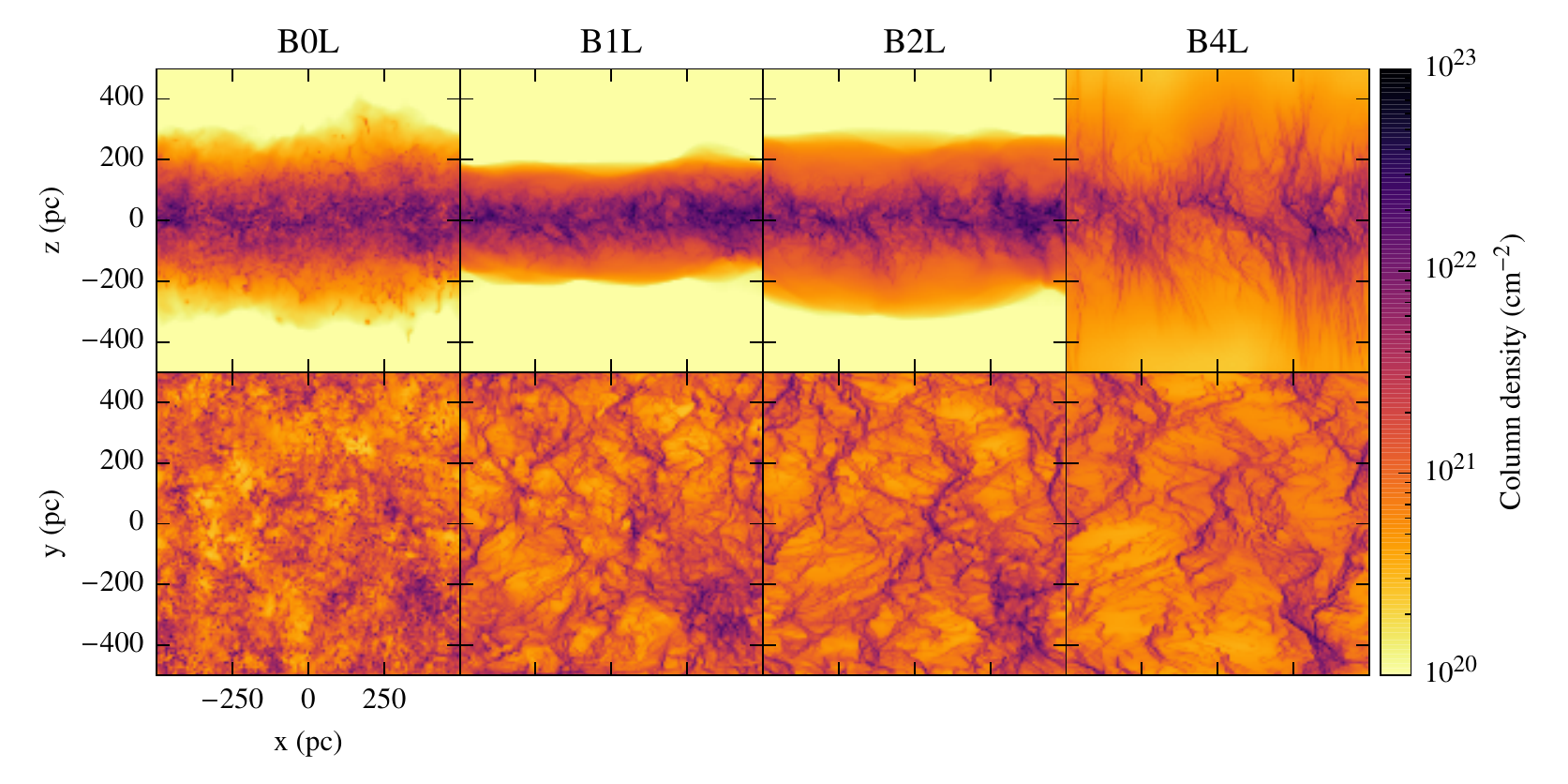}\\
            \includegraphics[width=15cm]{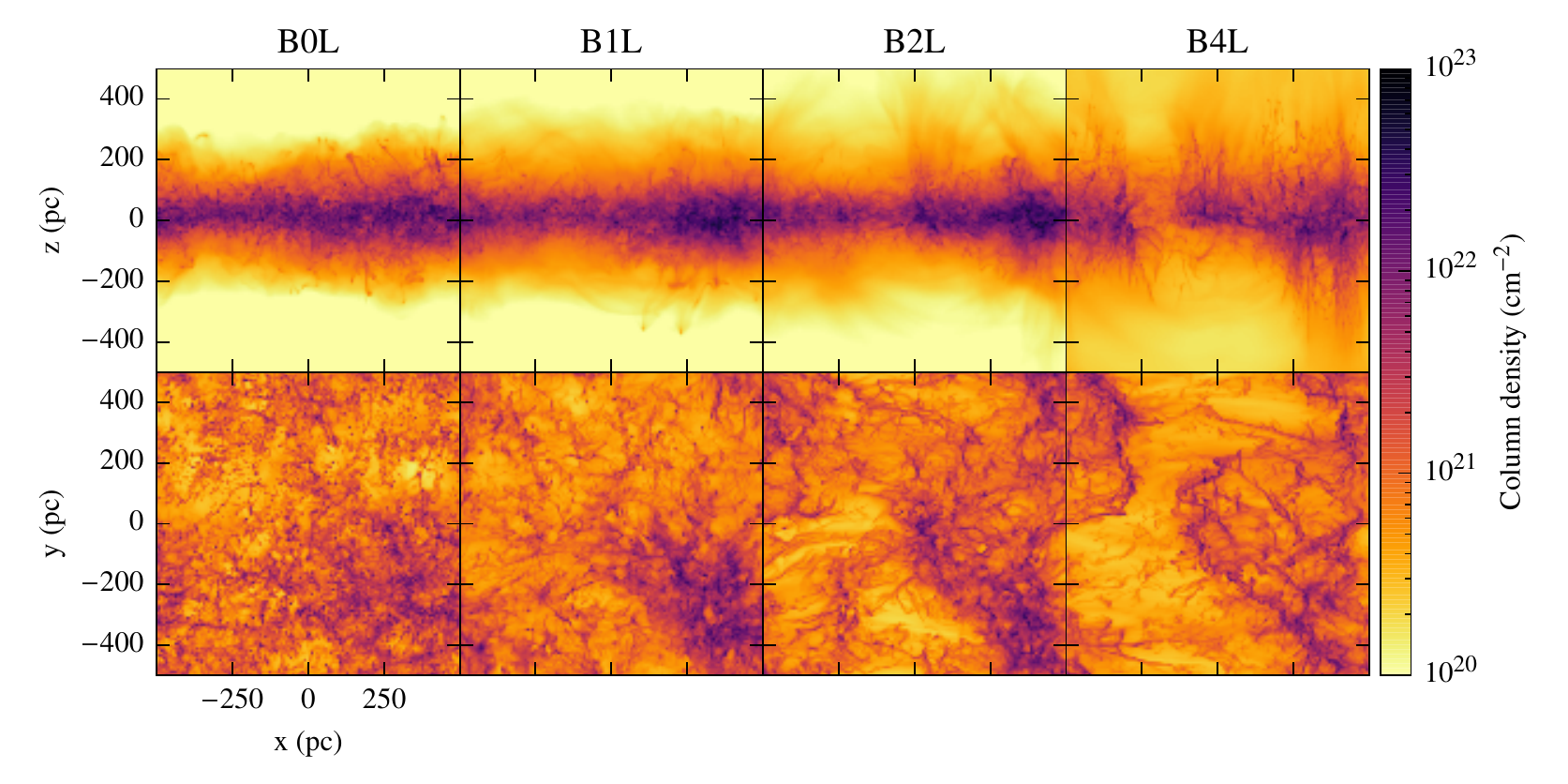}\\
            \includegraphics[width=15cm]{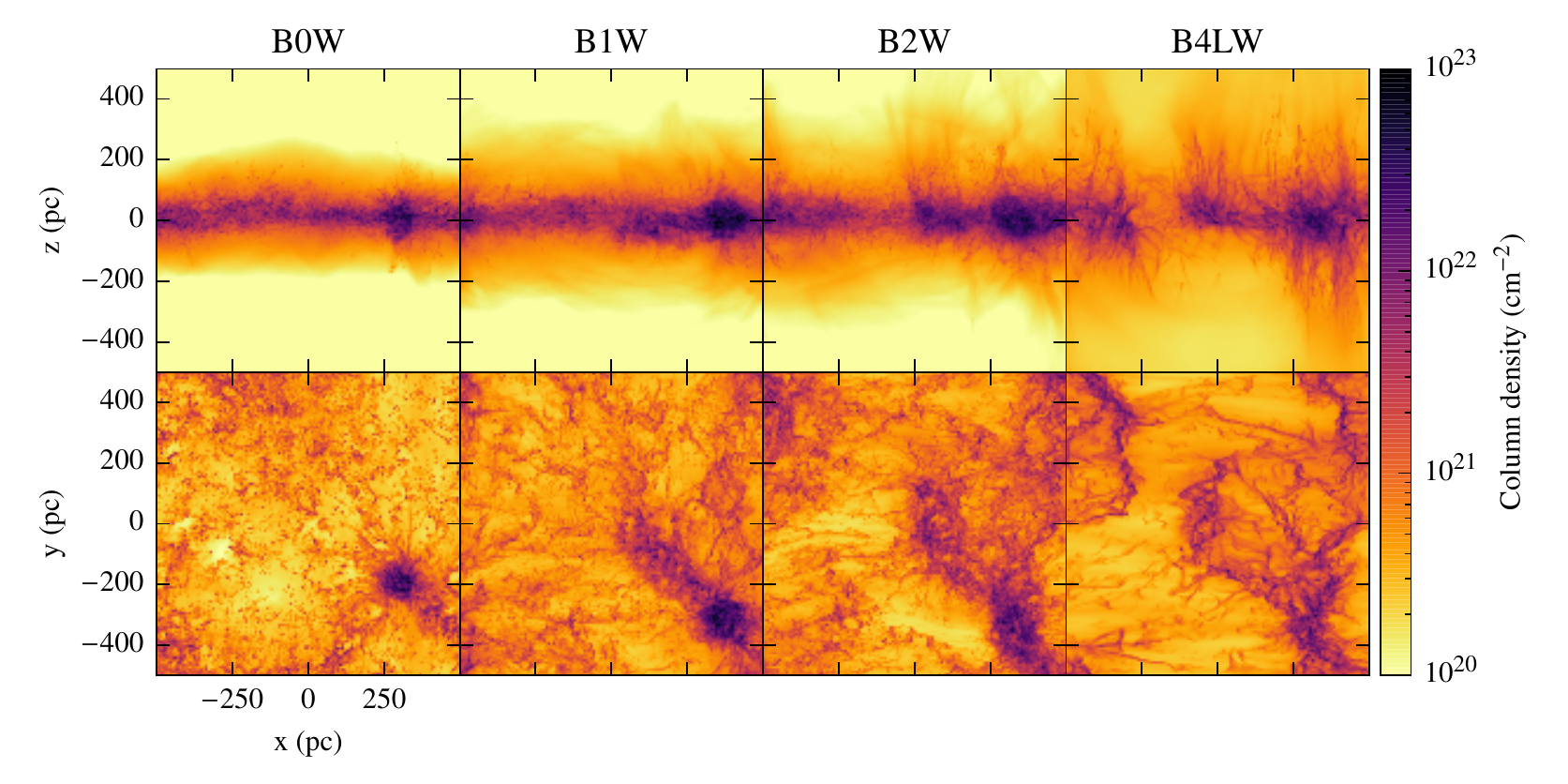}\\
         \end{center}
      \caption{Column density maps for the various runs.
      \emph{From left to right:} runs B0, B1, B2 and B4 (increasing initial
      magnetic field).
      \emph{From top to bottom:} Fiducial $p_{SN} = 4 \cdot 10^{43}\ \mathrm{g\
      cm / s}$ at $40$ and $80\ \mathrm{Myr}$, reduced $p_{SN} = 10^{43}\ \mathrm{g\
      cm / s}$ at $80\ \mathrm{Myr}$.
      \emph{First row:} edge-on view.
      \emph{Second row:} face-on view.
      }%
      \label{fig:rt}
      \end{figure*}

%__________________________________________________________________

\section{Numerical setup}\label{sect-num}

   \subsection{Numerical code and resolution}

      We run our simulations with the RAMSES code \citep{Teyssier02, Fromang06},
      an adaptive mesh code using a Godunov scheme and a constrained transport
      method to solve the MHD equations, therefore ensuring the nullity of the
      magnetic field divergence. We use a $512^3$ or $1024^3$ grid with no
      adaptive mesh refinement in order to get consistent power spectra: the
      interplay between an adaptive mesh and numerical diffusivity may introduce
      non-trivial biases into the spectra on a large range of scales because of variable 
      spatial resolution. This
      uniform grid has a cell size of about $2$ or $1\ \mathrm{pc}$. 

      Because of the small time steps induced by high velocities and temperatures generated by supernova explosions, 
      these simulations are quite demanding and have required altogether about 15 millions of 
      CPU hours on a BlueGene supercomputer. 

%__________________________________________________________________

      \begin{figure}
         \begin{center}
            \includegraphics[width=8cm]{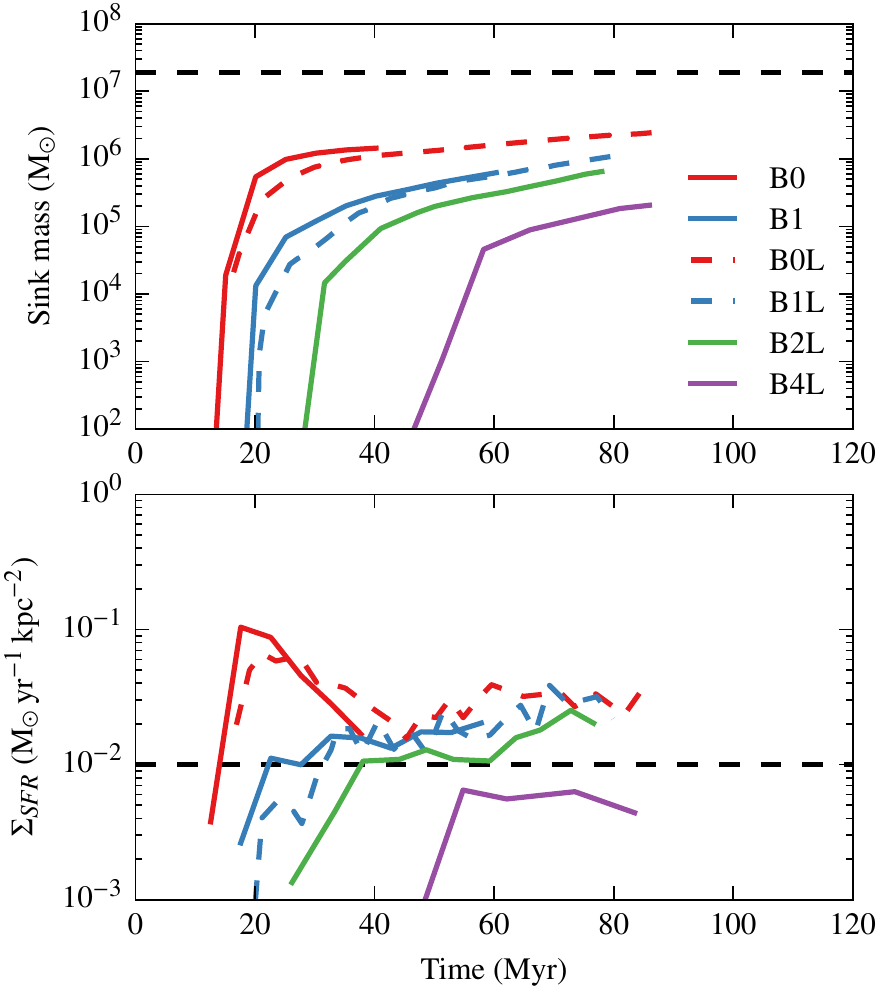}\\[1ex]
            \includegraphics[width=8cm]{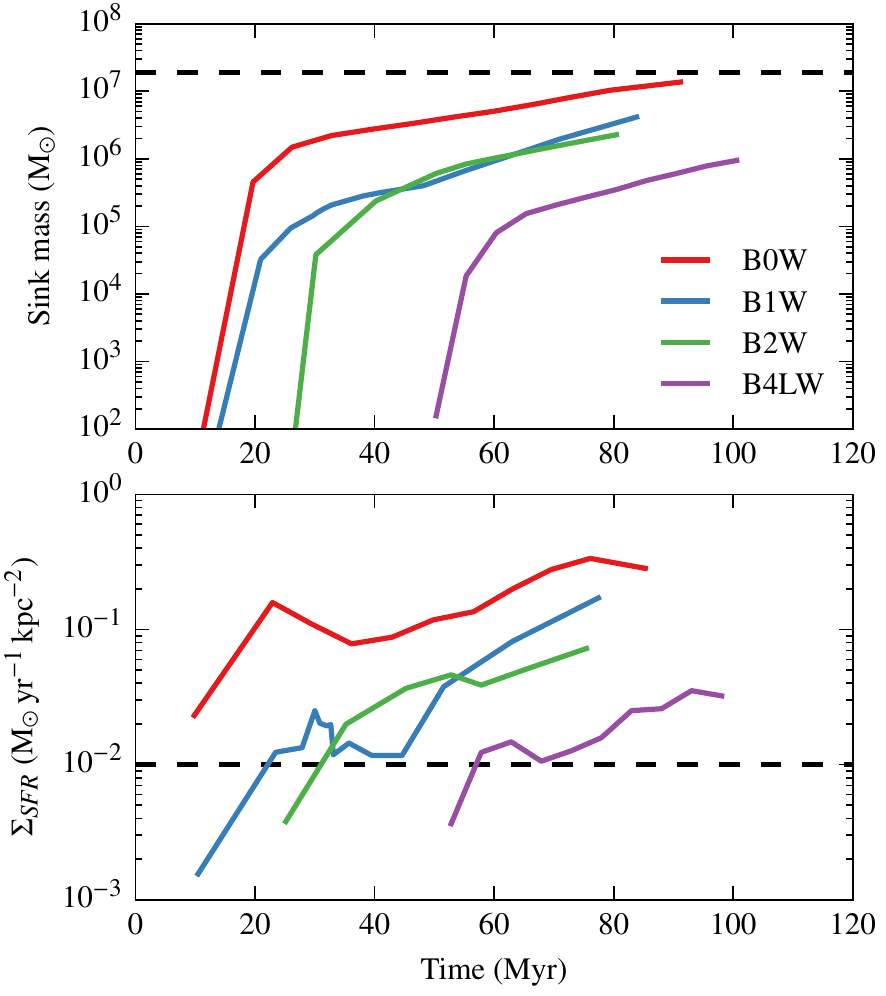}
         \end{center}
      \caption{Mass in sinks (\emph{first row}) and estimated star formation rate
      (\emph{second row}) as a function of time. The dotted line for the sink
      mass corresponds to the initial gas mass in the simulation box. The dotted
      line for the star formation rate corresponds to the typical observational value 
       \citep[e.g.][]{kennicut2007} given our column density.
      \emph{Top panels:} strong feedback.
      \emph{Bottom panels:} weak feedback.
      }\label{fig:msink_all}
      \end{figure}

   \subsection{Physical processes}

      Our simulations include various physical processes known to be important
      in molecular clouds. We solve the ideal magnetohydrodynamics (MHD)
      equations with self-gravity and take into account the cooling and heating
      processes relevant to the ISM. We also add an analytical gravity profile
      accounting for the distribution of stars and dark matter. The
      corresponding gravitational potential is \citep{Kuijken89b}:
      \begin{equation}
          \phi_{ext}(z) = a_1 \left(\sqrt{z^2 + z_0^2} - z_0\right) + a_2 \frac{z^2}{2},%
          \label{eq:ext-potential}
      \end{equation}
      with $a_1 = 1.42 \times 10^{-3}\ \mathrm{kpc}\ \mathrm{Myr}^{-2}$, $a_2 = 5.49 \times
      10^{-4}\ \mathrm{Myr}^{-2}$ and $z_0 = 180\ \mathrm{pc}$, as used by \citet{Joung06}.

      The equations we solve are
      \begin{align}
         \partial_t \rho + \vec{\nabla} \cdot \left( \rho \vec{v} \right) &= 0, \\
         \partial_t \left( \rho \vec{v} \right)
            + \vec{\nabla} \cdot \left( \rho \vec{v} \otimes \vec{v}
               + \left( P + \frac{B^2}{8\pi} \right) \tens{I}
               - \frac{\vec{B} \otimes \vec{B}}{4\pi} \right)
            &= -\rho\vec{\nabla} \Phi, \label{eq:mhd-momentum}\\
         \partial_t E
            + \vec{\nabla} \cdot \left(\left( E + P + \frac{B^2}{8\pi} \right) \vec{v}
               - \frac{1}{4\pi} \left( \vec{v} \cdot \vec{B} \right) \vec{B} \right)
            &= -\rho \vec{v} \cdot \vec{\nabla} \Phi - \rho\mathcal{L}, \\
         \partial_t \vec{B} + \vec{\nabla} \cdot \left( \vec{v} \otimes \vec{B}
            - \vec{B} \otimes \vec{v} \right) &= 0, \\
         \Delta \phi - 4\pi G \rho &= 0,\label{eq:mhd-poisson}
      \end{align}
      with $\rho$, $\vec{v}$, $P$, $\vec{B}$, $\Phi$, and $E$ respectively being
      the density, velocity, pressure, magnetic field, total gravitational
      potential, and total (kinetic plus thermal plus magnetic) energy. The loss
      function $\mathcal{L}$, includes UV heating and a cooling function with
      the same low-temperature part as in \citet{Audit05} and the
      high-temperature part based on \citet{Sutherland93}, resulting in a
      function similar to the one used in \citet{Joung06}. The gravitational
      potential $\Phi$ has two terms as stated before: the one due to stars and
      dark matter $\phi_{ext}$, and the one due to the gas itself $\phi$, hence
      $\Phi = \phi + \phi_{ext}$.

%__________________________________________________________________

   \subsection{Initial and boundary conditions}

      We initialize our simulations with a stratified disc: we use a Gaussian
      density profile:
      \begin{equation}
      n(z) = n_0 \exp \left[ - \frac{1}{2} \left( \frac{z}{z_0}  \right)^2 \right],
      \end{equation}
      where $n_0 = 1.5\ \mathrm{cm^{-3}}$ and $z_0 = 150\ \mathrm{pc}$. 
      This leads to a total column density, $\Sigma$, through the disc that is equal to 
      $\sqrt{2 \pi} \rho_0 z_0 $ where  $\rho_0=m_p n_0$ and $m_p$ is the mean mass per particle. 
      We get $\Sigma = 4 \times 10^{-3}$ g cm$^{-2} = 19.1$ M$_\odot$ pc$^{-2}$.

      For convenience, we define the scale height of the gas $H_0 =
      \Sigma / (2 \rho_0) = \sqrt{\pi / 2} z_0$. 

      The
      temperature is set to an usual WNM temperature, around $8000\ \mathrm{K}$.
      In order to prevent this disc from collapsing, an initial turbulent
      velocity field is generated with a RMS dispersion of $5\ \mathrm{km / s}$
      and a Kolmogorov \citep{Kolmogorov41} power spectrum with random phase. We
      add an initial horizontal magnetic field:
      \begin{equation}
      B_x(z) = B_0 \exp \left[ - \frac{1}{2} \left( \frac{z}{z_0} \right)^2 \right],
      \end{equation}
      with $B_0 \simeq 0, 3, 6 \text{ or } 12\ \mathrm{\mu G}$ for different runs
      (see Table~\ref{tab:runs}).  Note that our setup is not designed to provide a 
        detailed equilibrium along the z-axis since the latter must result from 
      a self-consistent feedback and star formation cycle. \,

       The boundary conditions are periodic in the x and y-directions and outflowing vanishing 
      gradient along the z-axis, that is to say the gas can leave the box but cannot enter.

%__________________________________________________________________

   \subsection{Supernova feedback}
   \label{sn_feed}

      The feedback scheme used for this set of simulations is similar to the
      fiducial run (C1) in \citet{Hennebelle14}.  We use Lagrangian sink
      particles \citep{Krumholz04, federrath2010, Bleuler2014} to track star-forming regions. 
      The specific implementation of \citet{Bleuler2014} uses a reconstruction of the 
      clumps on top of which sinks are placed if the clumps is gravitationally 
      unstable and collapsing.
      When such a
      sink particle accretes more than $120\ \mathrm{M_\odot}$ of gas, we assume
      a massive star has been formed, and we trigger a supernova within a $10\
      \mathrm{pc}$ radius around the sink particle, the exact position being determined
      randomly. This is done by injecting
      a few $10^{43}\ \mathrm{g\ cm / s}$ \citep{Iffrig15a} of radial
      momentum (later denoted $p_{SN}$, see also Table~\ref{tab:runs}) in a
      sphere (a few computational cells in radius) around the chosen location.
      The thermal energy is however saturated to a corresponding temperature of
      $10^5\ \mathrm{K}$ to prevent very high sound speeds outside the galactic
      plane, as such speeds enforce a very small time step, and such a run would
      be computationally too expensive (the number of required time steps is an
      order of magnitude higher). 
      In the same way, we also limit the velocity produced by the supernovae 
      to a maximum value of 200 km s$^{-1}$. We have tested  the influence of this threshold 
      by increasing its value and find that it does not have a strong influence. 
      These approximations give reasonable results
      for the gas in the disc, but the low-Mach outflowing gas is not treated
      correctly essentially because the hot phase produced by supernova explosions, 
      which is largely responsible of gas expulsion, is absent from these simulations. 

      Let us remind that the exact way supernova feedback is implemented has been found 
      to have a drastic influence on the results \citep{Hennebelle14,gatto2015}. In particular 
      if the supernovae are not sufficiently correlated with the dense gas, they do not 
      exert any substantial influence on the star forming gas and this leads to very high SFR. 
      Since treating precisely the dense gas and the supernova correlation implies resolving 
      not only the star formation well but also following the detailed star trajectories, this 
      constitutes a very difficult task, that renders prescriptions like the one we are using unavoidable.

      More generally, other sources of feedback such as HII radiation and stellar winds should be 
      considered as well \citep{walch2012,dale2013,dale2014,geen2015,geen2016}.

   \subsection{Runs performed}
      The
      different runs are summarized in Table~\ref{tab:runs}.

      \begin{table}
         \begin{center}
            \begin{tabular}{lccc}
               \hline\hline
               Name & {$B_0$ ($\mathrm{\mu G}$)} & {$p_{SN}$ ($\mathrm{g\ cm\ s^{-1}}$)} & Resolution \\
               \hline
               B0L  & 0                          & $4 \cdot 10^{43}$                 & $512^3$ \\
               B1L  & 3                        & $4 \cdot 10^{43}$                 & $512^3$ \\
               B2L  & 6                          & $4 \cdot 10^{43}$                 & $512^3$ \\
               B4L  & 12                         & $4 \cdot 10^{43}$                 & $512^3$ \\
               \hline
               B0   & 0                          & $4 \cdot 10^{43}$                 & $1024^3$ \\
               B1   & 3                        & $4 \cdot 10^{43}$                 & $1024^3$ \\
               \hline
               B0W  & 0                          & $10^{43}$                         & $1024^3$ \\
               B1W  & 3                        & $10^{43}$                         & $1024^3$ \\
               B2W  & 6                          & $10^{43}$                         & $1024^3$ \\
               B4LW & 12                         & $10^{43}$                         & $512^3$ \\
               \hline
            \end{tabular}
         \end{center}
         \caption{Summary of the runs. The resolution corresponds to the number
         of cells used to cover the $1\ \mathrm{kpc}$ cube.}\label{tab:runs}
      \end{table}

      To understand the influence of the magnetic field we consider four values of the initial 
      magnetic intensities while to study the impact of feedback we perform runs with two values 
      of $p_{SN}$, the momentum injected by supernovae. 
      The first value, $4 \times 10^{43}$ g cm$^{-1}$, corresponds to the typical momentum 
      thought to be injected in the ISM by supernovae and we will refer to it as ``standard''. 
      We will refer to the second value of $10^{43}$ g cm$^{-1}$ as ``weak''. 
      By combining low resolution (labeled ``L'') and high resolution runs, we can check for numerical 
      convergence. All simulations, except B0, have been run up to 80 Myr.

%__________________________________________________________________

\section{Global properties}

%__________________________________________________________________
In this section we give a general description of the simulations and 
 quantitative estimate of the SFR and disc scale height.

   \subsection{Qualitative description}

      Figure~\ref{fig:rt} shows column density maps for the various runs.
      Apart from the highest resolution, the general features are similar 
      to the ones of the lower resolution runs presented in \citet{Hennebelle14} and also
      seen in other works \citep[e.g.][]{kim+2011,walch2015,Kim15a,Kim15b}.
      In particular, a stratified multi-phase disc developed. As expected its density 
      profile varies with magnetic field and feedback strength. By comparing the snapshots 
      at 40 and 80 Myr, it is clear that the disc profile significantly evolves with time
      and becomes progressively thinner, a point that is closely analyzed below.

      The most visible effect of the variation of the initial magnetic field
      intensity is the change  of the disc thickness at time 40 Myr.  While this is expected because 
      qualitatively a stronger magnetic
      field entails more magnetic support, and therefore the disc should be thicker, 
      we will see later that quantitatively this may not be so simple.
      Another effect of   the magnetic field is also that it reduces the
      star formation rate and delays the beginning of the star-forming phase.
      This explains the thicker disc in the B0L case, where star formation starts
      earlier, which in turn implies more feedback from supernovae. Visually,
      one can also notice that the medium is rather filamentary in the
      magnetized run, whereas it looks clumpy in the hydrodynamical case \citep{hennebelle2013}.

      Comparing the middle and bottom panels of Fig.~\ref{fig:rt}, the effect
      of reduced momentum input is relatively clear: there is less support
      against gravity, hence a thinner disc. The horizontal structure is also
      affected, resulting in a dense structure surrounded by diffuse medium,
      whereas a stronger supernova feedback disperses the high-density clumps
      more easily, resulting in lower column density contrasts. 
      Moreover we see that  a very dense region has developed in runs B0W 
      (at $x \simeq 250$ pc and $y \simeq -200$ pc) and B1W 
      and is probably developing in runs B2W and B4W. This shows that 
      global collapse is going on because feedback is too weak to prevent it.

      \begin{figure}
         \begin{center}
            \includegraphics[width=8cm]{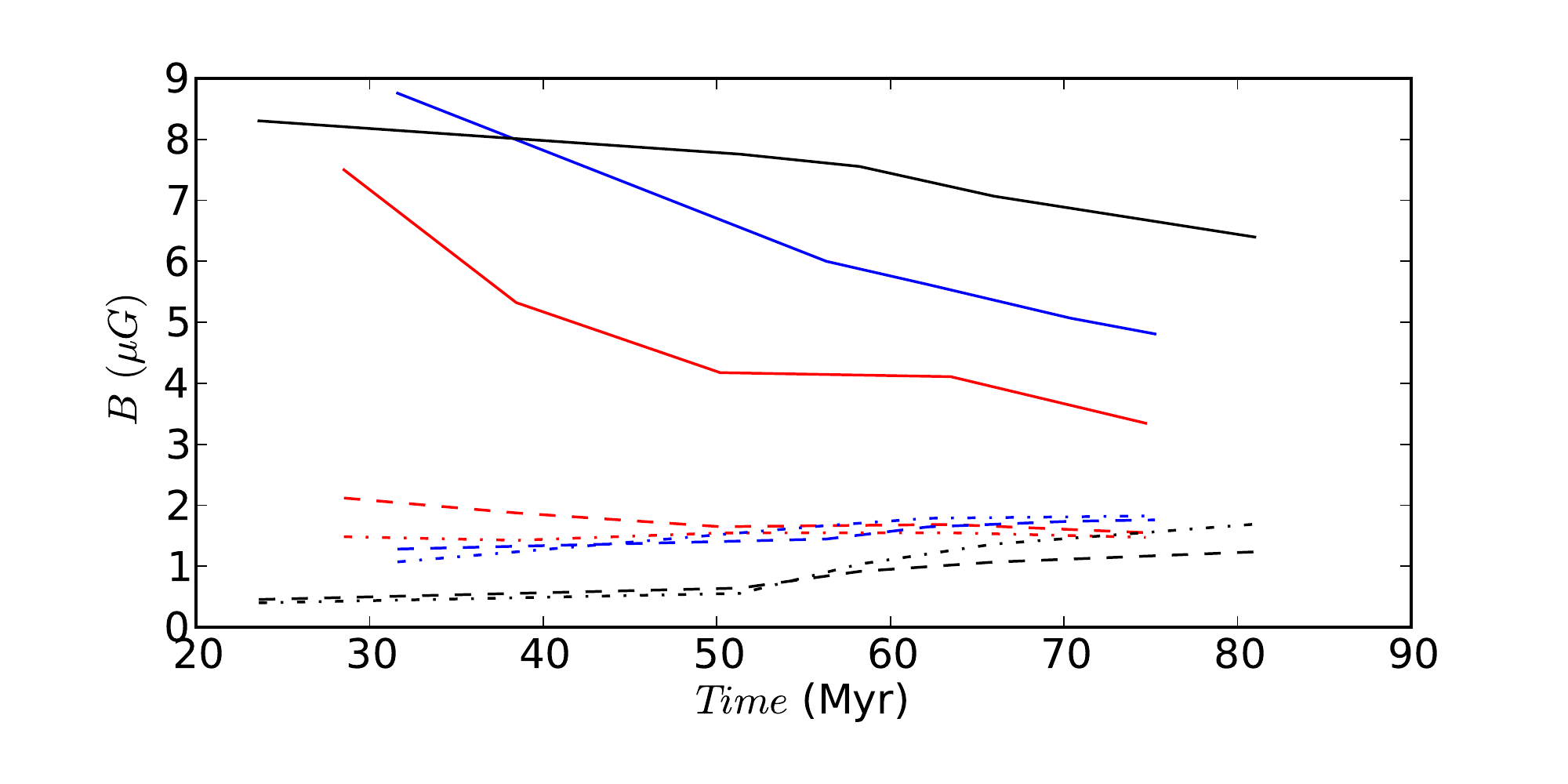}
         \end{center}
      \caption{Magnetic field in the equatorial plane as a function of time for 
        the three runs B1L, B2L and B4L. Solid line is the x-component while dotted 
        and dot-dashed are the y and z-components respectively. As can be seen 
        the x-component decreases with time which indicates that the initial 
        magnetic field gets expelled from the galactic plane. 
      }%
      \label{fig:mag_evol}
      \end{figure}

%__________________________________________________________________

   \subsection{Star formation history}\label{sect:sfr}

      The star formation rate is  estimated by following the mass
      accreted by the sink particles as a function of time.
      Figure~\ref{fig:msink_all} displays the total mass and the SFR (obtained 
      by taking the time derivative of the sink mass with respect to time). 
      For reference the corresponding estimate is given for a typical Milky-Way type galaxy. 
       Typical 
      SFR in the simulations are on the order of a few 10$^{-2}$ M$_\odot$ yr$^{-1}$ kpc$^{-2}$. Given 
      the total column density of about 19 M$_\odot$  pc$^{-2}$, these SFR appears to be 
      comparable to observed values \citep[e.g.][]{kennicut1998,kennicut2007,bigiel2008}.

\begin{figure}
         \begin{center}
            \includegraphics[width=8cm]{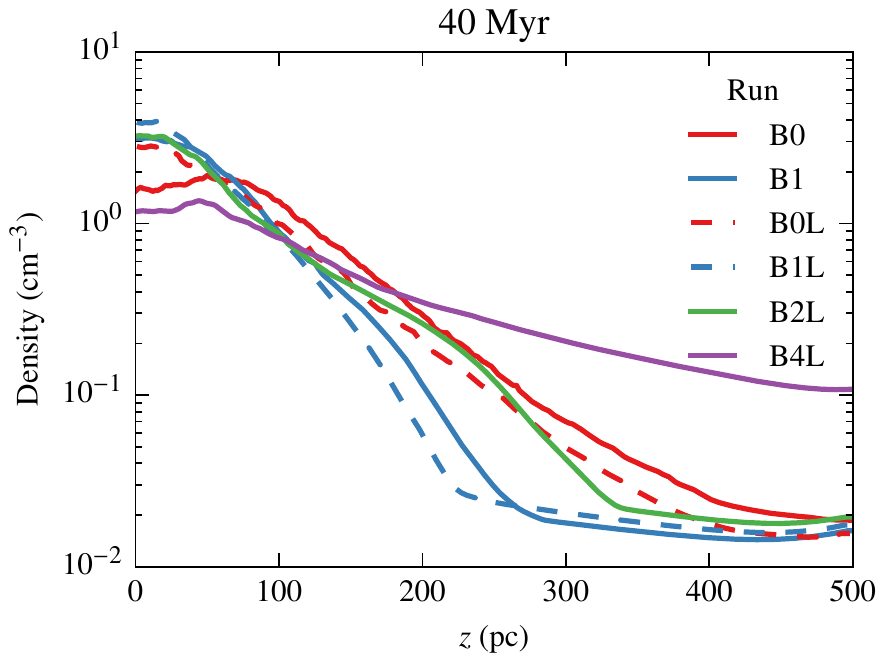}\\[1ex]
            \includegraphics[width=8cm]{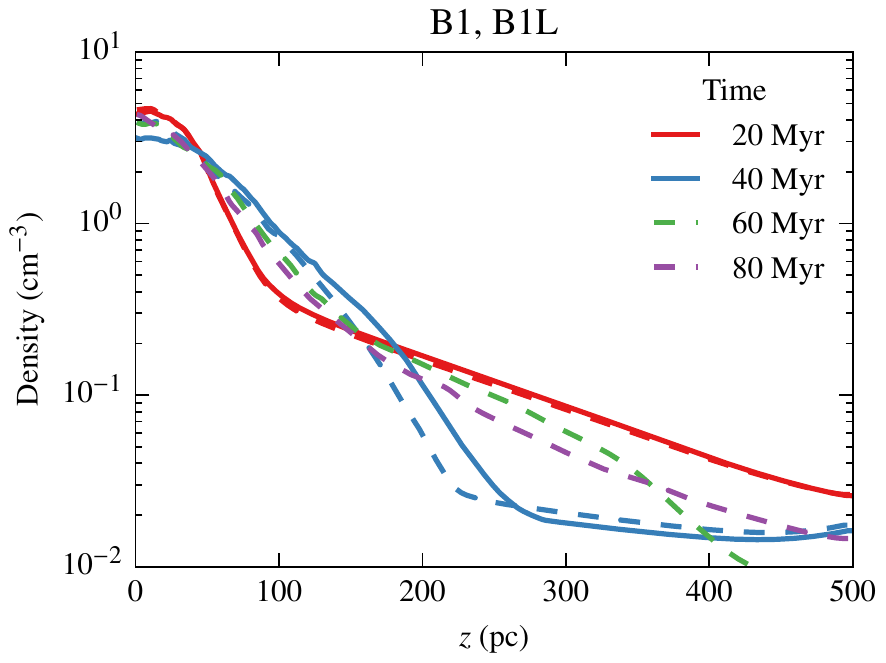}%\\[1ex]
         \end{center}
      \caption{Vertical density profiles.
         \emph{Top:} Disc profile for the different runs with strong feedback,
         at $40\ \mathrm{Myr}$.
         \emph{Bottom:} Evolution with time in the B1 and B1L runs. The dashed
         lines are the lower-resolution profiles.
%         \emph{Bottom:} Evolution with time in the B1W run.
      }\label{fig:rho_profiles}
      \end{figure}

      As expected, Fig.~\ref{fig:msink_all} shows that the SFR is reduced
      in the presence of  a stronger initial magnetic field by a factor up to 4 in the 
      standard feedback case and 10 in the weak feedback one. 
      Another effect of  stronger
      fields is that star formation is delayed by roughly $30\ \mathrm{Myr}$
      between runs B1L and B4L. While these effects are substantial, one may wonder 
      that they could be even stronger and that star formation could, in principle  
      be almost entirely quenched. This is because for a 10 $\mu G$ magnetic field, 
      the Alfv\'en speed within the WNM is about 20 km s$^{-1}$ and therefore 
      2-3 times larger than the sound speed that is equal to about 7 km s$^{-1}$. 
      Therefore the magnetic Jeans mass is 8-20 times the thermal Jeans mass. 
      To investigate this, Fig.~\ref{fig:mag_evol} displays the mean value of the 
      magnetic intensity in the equatorial plane for the runs B1L, B2L and B4L. 
      As can be seen for the runs B1L and B2L the magnetic intensity at 30 Myr is above 
      their initial values (respectively about 2.5 and 5 $\mu G$), this is due to the initial 
      contraction of the galactic plane and the magnetic flux freezing, that tends to amplify 
      the magnetic field. 
      Clearly the $B_x$ component decreases with time and comes close to the initial field
      intensity for run B1L and B2L. For run B4L, it falls below (almost a factor of 2) the
      initial intensity. This indicates that some magnetic flux is expelled from the galactic 
      plane and does not remain frozen into the gas.  Let us stress that the 
      galactic disc is clearly contracting after 40 Myr and nearly stationary between times 60 and 80 Myr
      (see bottom panel of Fig.~\ref{fig:rho_profiles}), 
      therefore the flux decrease is not a consequence of global expansion of the disc. 
      It is likely a consequence of the 
      turbulence, which triggers reconnection and magnetic diffusivity a process largely observed in 
      other contexts \citep{lazarian1999,joos2013}.
      On the contrary, in the three cases  the $B_y$ and $B_z$ components are amplified likely because of turbulence,
      to values on the order of 1-2 $\mu G$.
      To conclude, magnetic field does reduce the SFR but its impact is mitigated by turbulent magnetic flux leakage.
      We stress that since shear is not included in our study, we may neglect or underestimate the generation 
      of magnetic field.

      In the weak feedback runs (bottom panels of Fig.~\ref{fig:msink_all}), the SFR is higher than for the standard feedback case
      by a factor 3-10 (depending of time and runs). It is interesting to note that contrary to the standard 
      feedback case, where a plateau is reached, the SFR keeps increasing with time indicating that no stationary behaviour has been reached. 
      This is likely because the feedback is too weak and therefore the galactic disc is undergoing 
      a catastrophic collapse as it is indeed suggested by Fig.~\ref{fig:rt}.

      We conclude that the scheme employed here is able to reproduce the observed SFR, which is a fundamental 
      property for galaxies. While the total amount of momentum delivered in the flow is roughly correct, 
      we caution that changing the correlation between supernovae and dense gas may alter 
      this conclusion. This however suggests that the present setup can be employed to study further 
      the flows and the dense structures arising in galactic discs.  

      \begin{figure}
         \begin{center}
            \includegraphics[width=8cm]{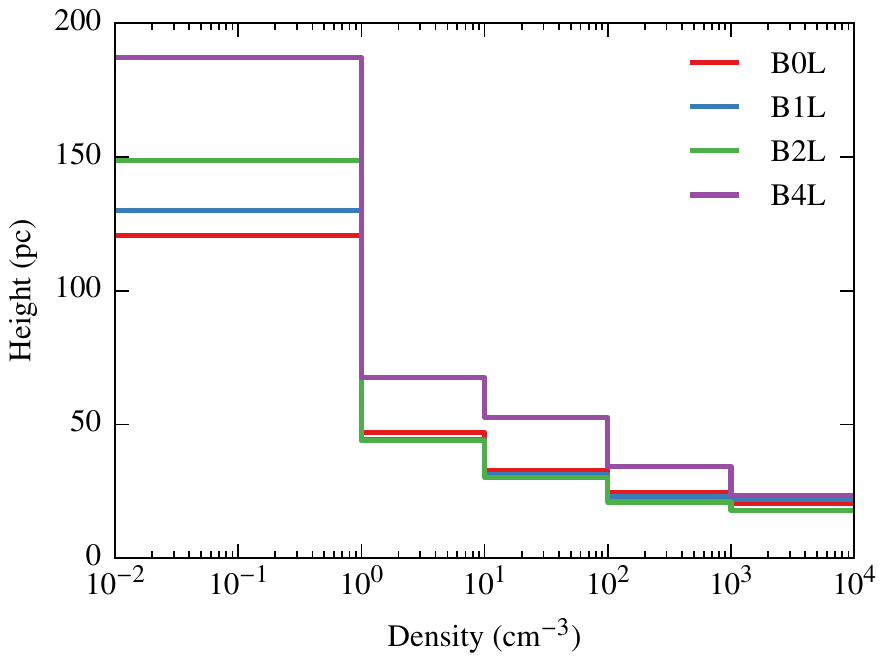}
         \end{center}
      \caption{Mean height as a function of gas density for runs
        B0L, B1L, B2L and B4L at time 60 Myr. 
              }%
      \label{fig:heights}
      \end{figure}

      \begin{figure}
         \begin{center}
            \includegraphics[width=8cm]{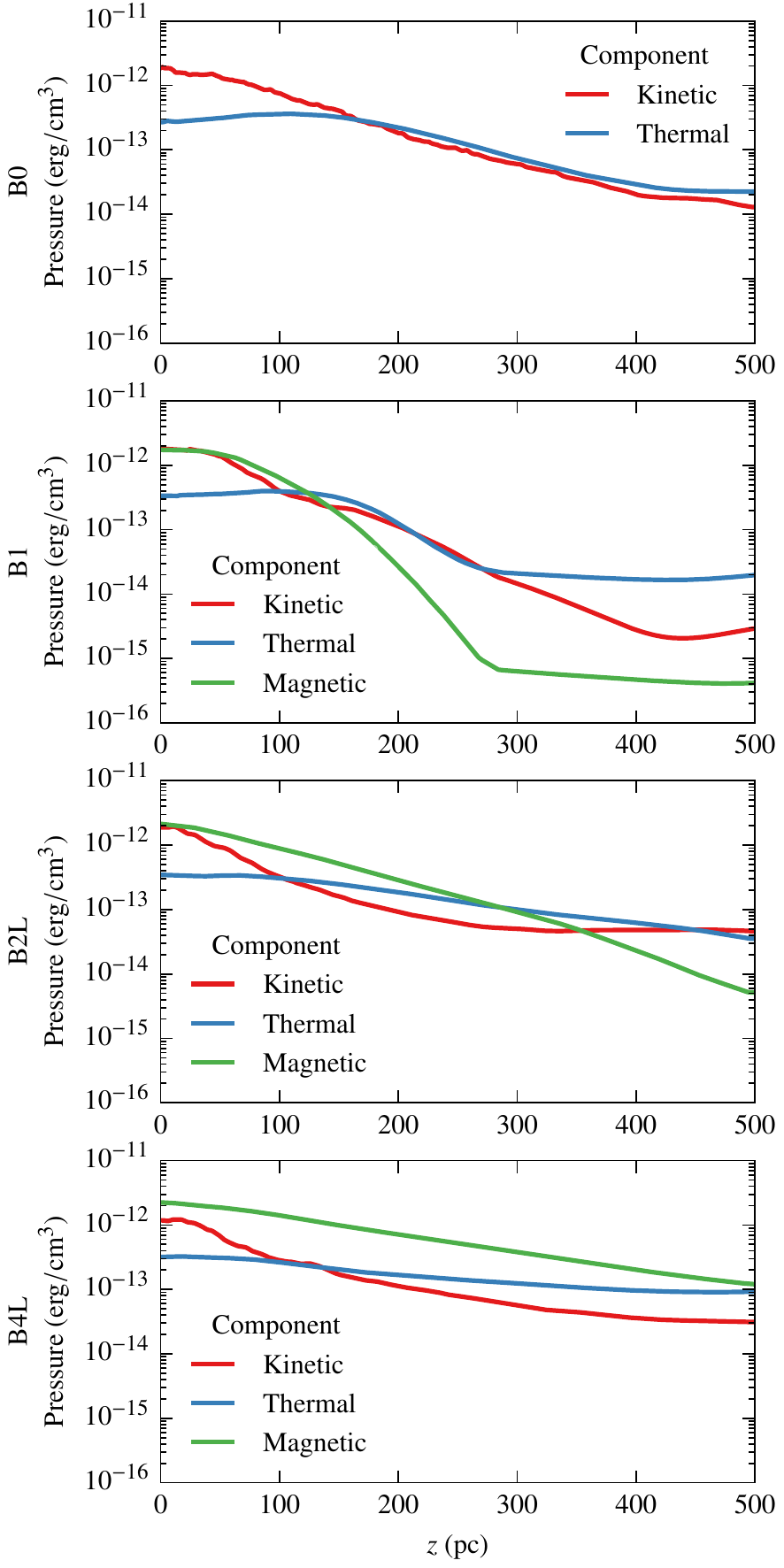}
         \end{center}
      \caption{Vertical kinetic, thermal and magnetic pressure profiles
      $25\ \mathrm{Myr}$ after the beginning of star formation.
      \emph{From top to bottom:} runs B0L, B1L, B2L and B4L.
      }\label{fig:p_profile_all}
      \end{figure}

      \begin{figure}
         \begin{center}
            \includegraphics[width=8cm]{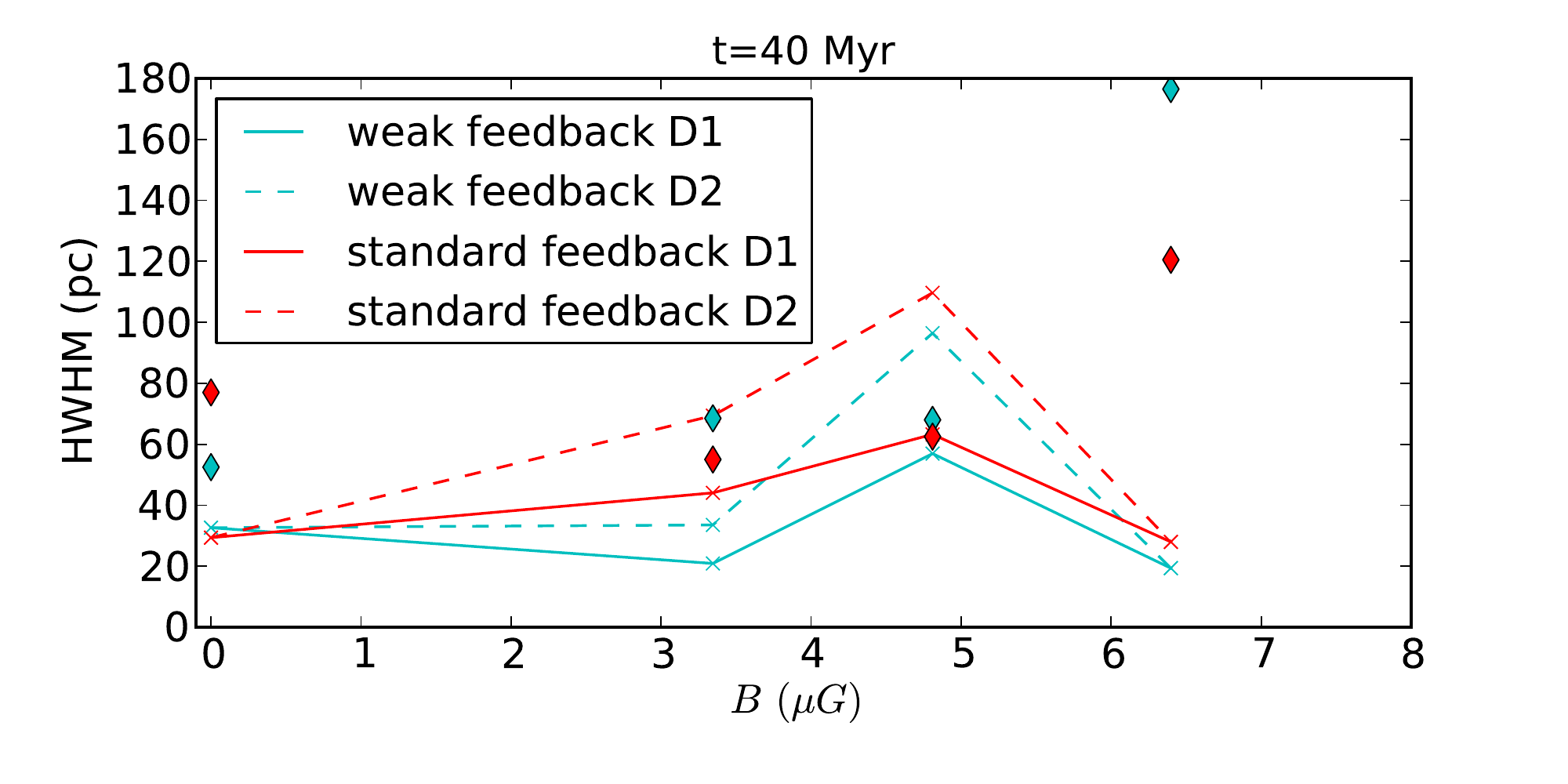}\\[1ex]
            \includegraphics[width=8cm]{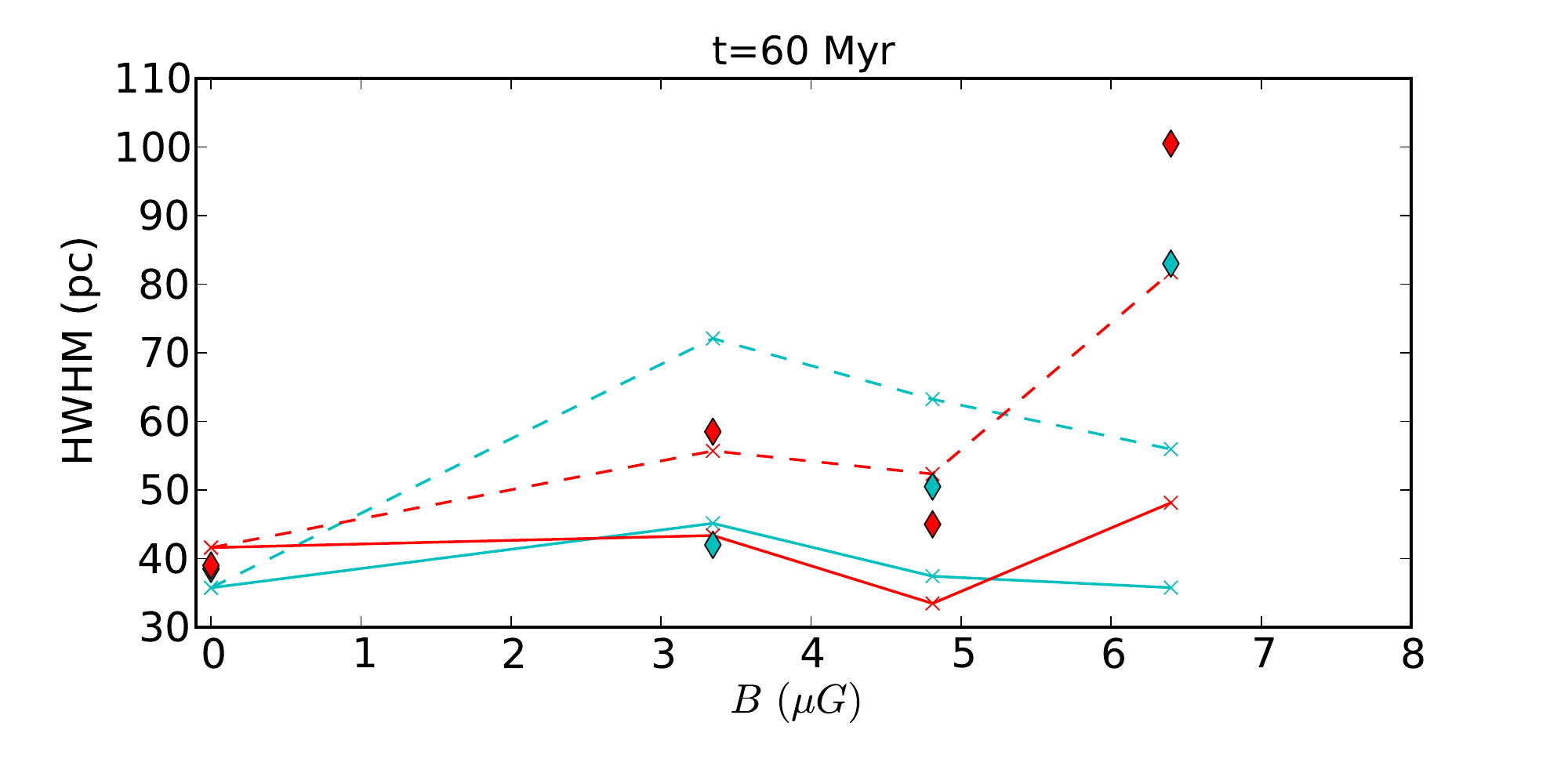}\\[1ex]
            \includegraphics[width=8cm]{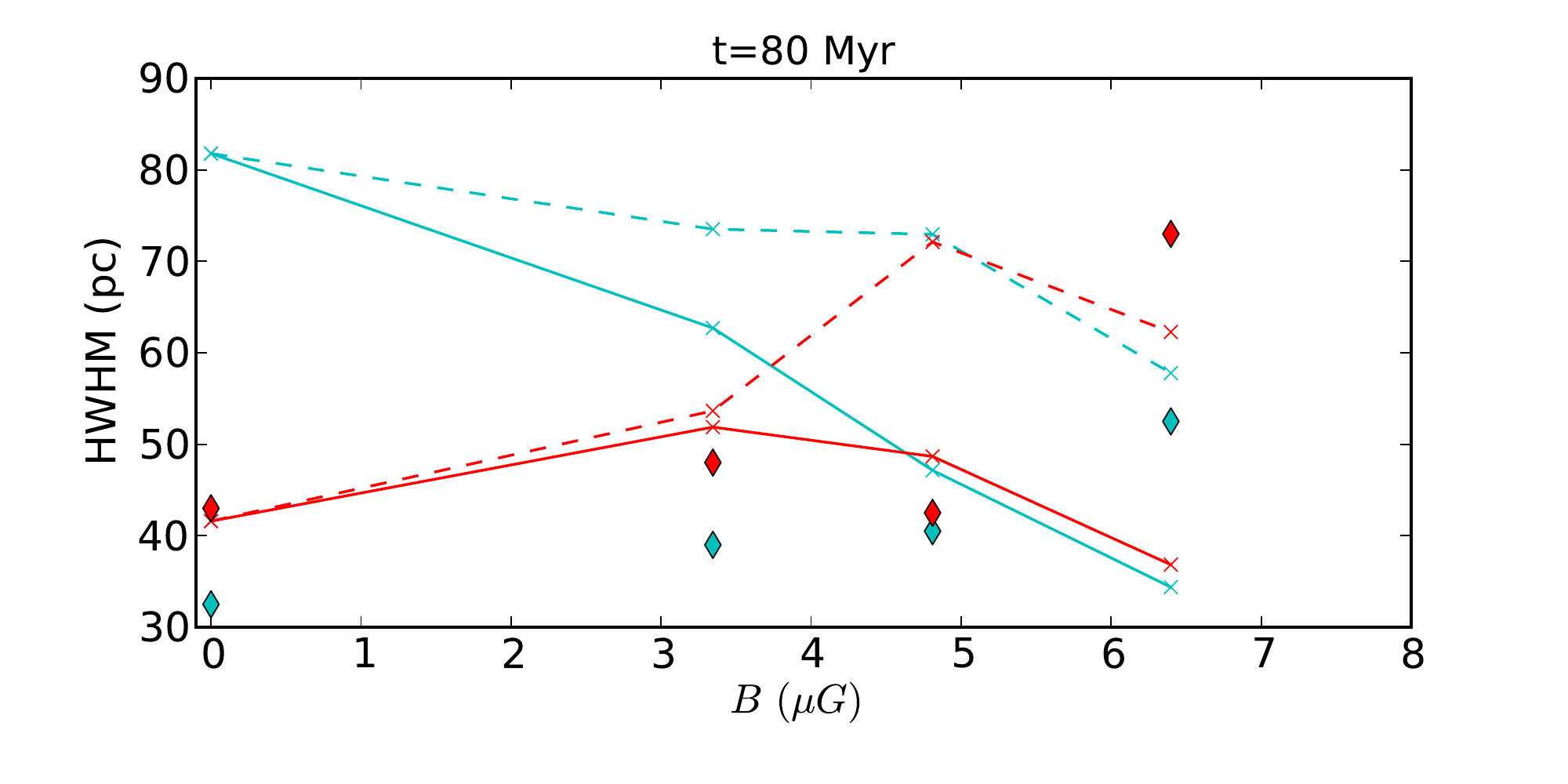}
         \end{center}
      \caption{Comparison between the half width at half maximum (HWHM) of the density profiles 
        measured in the simulations (diamonds) and the analytical models at time 40, 60 
        and 80 Myr.  Red lines
        represent the standard feedback case and cyan ones the weak feedback case.
        Solid lines stand for constant  dissipation while dashed ones 
        represent magnetically dependent dissipation (see text). 
      }\label{fig:model}
      \end{figure}

%__________________________________________________________________

%__________________________________________________________________

   \subsection{Vertical structure}\label{sect:height-model}

   \subsubsection{Density and pressure profiles}

      In order to study the vertical structure of the simulated galactic disc,
      we compute vertical profiles by averaging over horizontal slices.
      The resulting density profiles are shown in the top panel of 
      Fig.~\ref{fig:rho_profiles}.  The
      corresponding full widths at half maximum (FWHM) are respectively $160$,
      $110$, $110$, and $240\ \mathrm{pc}$ for runs B0L, B1L, B2L, and B4L at
      $40\ \mathrm{Myr}$. These values are comparable to the observed values for
      the Milky Way: $120\ \mathrm{pc}$ for the molecular gas and $230\
      \mathrm{pc}$ for the atomic gas \citep{Ferriere01}, although there is no
      distinction between those components in the simulations.  In an attempt to 
      better quantify the scale height in our simulations, we have calculated the mean height
      per density bin and displayed the corresponding result in Fig.~\ref{fig:heights}.
      Using the definition of \citep{Ferriere01} (see her Eq.~1), which is slightly different 
      from the one we use, and the values $H_m = 71-81\ \mathrm{pc}$ that she quotes, this corresponds for 
      the molecular gas to a mean height of about $40-45\ \mathrm{pc}$ and $100\ \mathrm{pc}$ for the HI. 
      These numbers are in reasonable agreement for the gas of densities between $10$ and $100\ \mathrm{cm^{-3}}$ 
       and less dense than $1\ \mathrm{cm^{-3}}$ respectively. They are about two times 
      larger for the gas denser than $100$ and $1\ \mathrm{cm^{-3}}$ respectively. It is therefore hard to conclude
      whether the disc is too narrow by a factor 2 or compatible with the observations as this would imply 
      a good description of the molecular gas in the simulation (as well as of the CO emission), which in particular
      likely requires a better spatial resolution. It is worth stressing that our simulations do not 
      include the cosmic rays, which likely contribute to support the galactic disc against gravity. 
      In particular \citet{girichidis+2016} have recently performed simulations which suggest that 
      this cosmic rays could indeed have a significant contribution, mainly because
      they dissipate less easily that the hot gas produced by supernova explosions. Definite 
      conclusions are however prevented given the difficulties to  accurately measure this scale height.

      The bottom panel of Fig.~\ref{fig:rho_profiles} shows the density profiles 
      at various time steps for run B1. As can be seen it follows a complex evolution. First 
      of all between $20$ and $40\ \mathrm{Myr}$, the disc starts expanding. This is a consequence
      of the increasing star formation between these two time-steps. Then the disc  
      contracts and finally reaches an equilibrium, as shown by the comparison between times
      60 and 80 Myr. The comparison between runs B1 and B1L shows that numerical 
      convergence seems to have been reached, at least for this particular aspect. 

      To get a better understanding of the vertical equilibrium, we show the various 
      pressures, namely thermal, kinetic and magnetic across the galactic disc and for
      the 4 simulations B0, B1, B2L and B4L about $25\ \mathrm{Myr}$ after star formation has begun.
      As found in other studies \citep[e.g.][]{Kim13,Hennebelle14}, we find that
      the thermal pressure is negligible while the kinetic one tends to be dominant. 
      We find that the magnetic pressure is typically comparable to the kinetic one
      except in the most magnetized case for which it dominates. While the values at $z=0$
      are relatively similar, the profiles present significant differences. In particular 
      the stronger the magnetic intensity, the smoother its distribution. For
      example while for the B1 run the magnetic pressure falls to very low values above $z=300$ pc,
      at this height it is still one tenth of its value at the origin for run B4L. This clearly shows the magnetic flux expulsion 
      discussed in the previous section (Fig.~\ref{fig:mag_evol}). In particular these profiles do not correlate well with the 
      density ones while flux freezing would imply strict proportionality.

      \begin{figure}
         \begin{center}
            \includegraphics[width=\hsize]{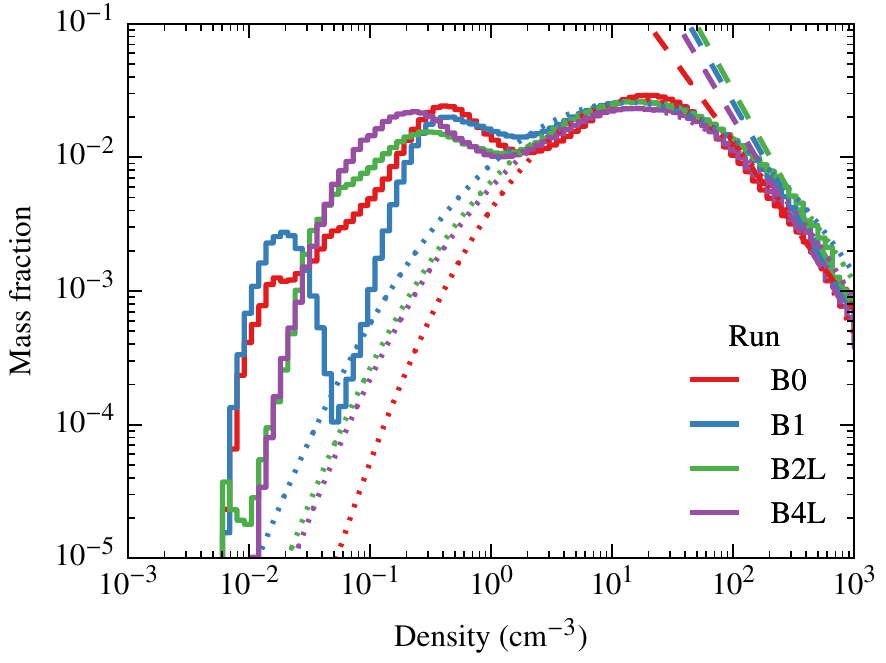}
            \includegraphics[width=\hsize]{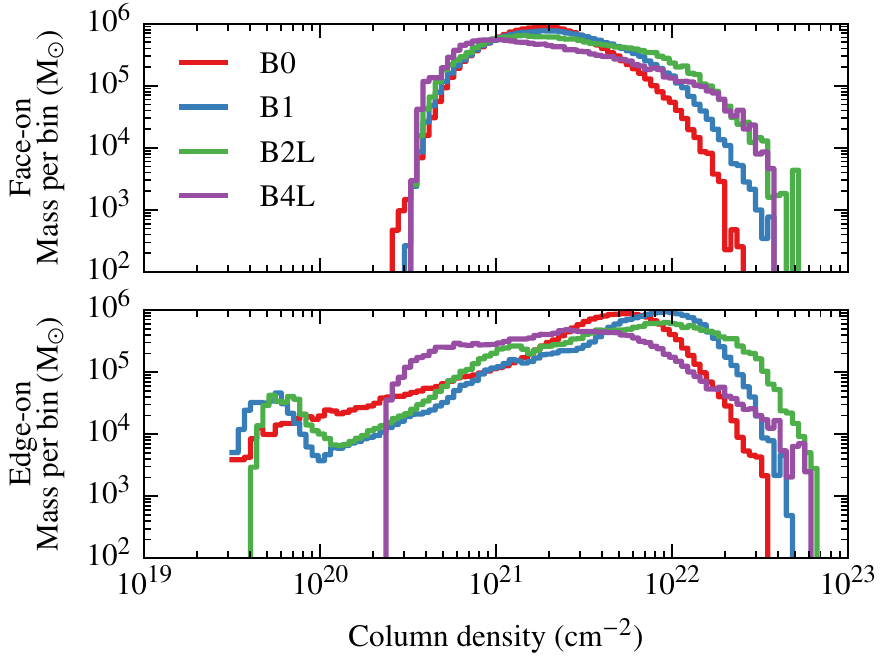}
         \end{center}
      \caption{Density (upper panel) and column density (lower pannels) distributions for the runs with strong feedback. The
      dotted parabolas are log-normal fits between $1$ and $100\
      \mathrm{cm^{-3}}$ and the dashed lines are tentative power-law fits above $100\
      \mathrm{cm^{-3}}$, with slopes $-1.1$, $-1.5$, $-1.6$, and $-1.5$ for
      runs B0L, B1L, B2L, and B4L respectively.
      }%
      \label{fig:n_hist}
      \end{figure}

         \begin{figure}
            \begin{center}
               \includegraphics[width=8cm]{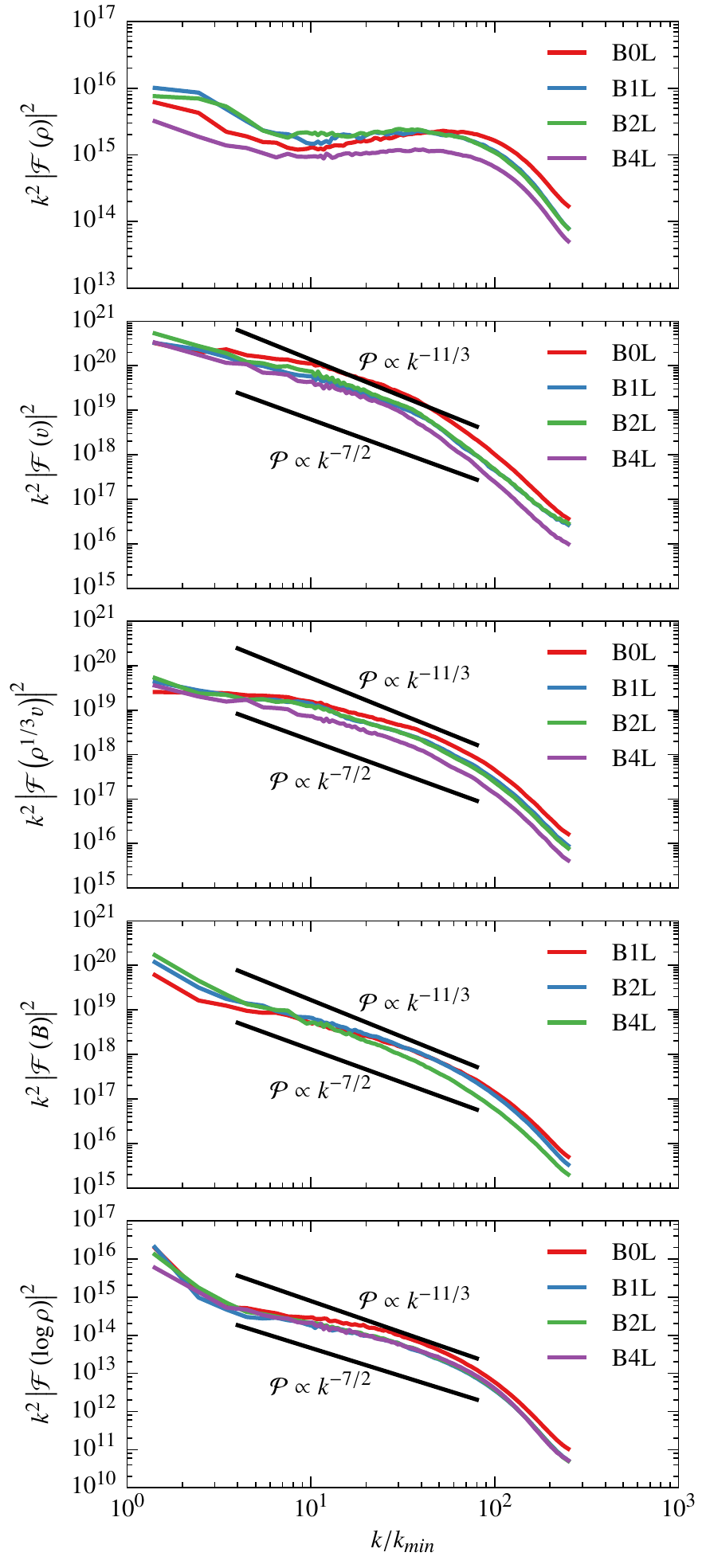}
            \end{center}
         \caption{Three-dimensional power spectra.
         \emph{From top to bottom:} density, velocity, density-weighted
         velocity, magnetic field.
         The spectra are multiplied by $k^2$ (such that the Kolmogorov scaling
         corresponds to a slope of $-5/3$)}%
         \label{fig:pspec_m_3d_s}
         \end{figure}

   \subsubsection{Analytical modelling}
      Although the various profiles display a great complexity, it is worth building an analytical model 
      to get a better quantitative insight.  
       For this purpose we assume that the disc is stationary. While this is obviously not the case 
      at early time, Figs.~\ref{fig:msink_all}-\ref{fig:rho_profiles} 
      show that this is a very reasonable assumption at time 60 and 
      80 Myr for the standard feedback case.
      We can derive an estimate for the disc thickness using the approach of
      \citet[][Sect.~2]{Kim15b}.  By averaging the momentum conservation
      equation~\eqref{eq:mhd-momentum} in time and horizontal directions
      \citep[see e.g.][]{Boulares90}, and integrating it between the mid-plane
      and some altitude $z_{max}$, where we assume the density to be negligible,
      we can write:
      \begin{equation}
          \rho_0 \sigma_z^2 - \Delta \Pi_{mag}  = \mathcal{W}_{self} +
          \mathcal{W}_{ext},%
          \label{eq:height-equil}
      \end{equation}
      where $\sigma_z$ is the vertical component of the (thermal and turbulent)
      velocity dispersion, $\Delta \Pi_{mag}$ is the difference of magnetic
      support between $z_{max}$ and the mid-plane, $\mathcal{W}_{self}$ is the weight of
      the gas disc in its own gravitational potential, and $\mathcal{W}_{ext}$
      is the weight of the gas disc in the external potential. Expressions for
      these quantities will be derived in the following paragraphs.

      In order to estimate $\sigma_z$, we assume that a stationary energy equilibrium 
      is established between on one hand the supernova energy injection and on the 
      other hand the turbulent energy dissipation, we write 
      \begin{eqnarray}
        {\Sigma \sigma^3   \over H } = \eta_{turb}  \dot{n}_{SN} \epsilon_{SN} E_{SN},
      \end{eqnarray}
      where $\dot{n}_{SN}$ is the number of supernova explosions per unit of time 
      that we assumed to be equal to $\Sigma _{SFR} / 120\ \mathrm{M_\odot}$ that is to say the 
      mass of stars produced per unit of time divided by the mass necessary to get 
      a massive star (assuming a canonical Salpeter IMF), $ E_{SN}$ is 
      the energy of the supernova taken to be equal to $10^{51}$ erg and $\epsilon_{SN}$
      is the fraction of this energy that is effectively communicated to the gas. As 
      a canonical value we will adopt $5\,\%$. Then finally $\eta_{turb}$ is a coefficient 
      of the order of 1 and that will be adjusted. It reflects our ignorance of the exact way 
      that turbulence is decaying, in particular since the galactic disc is stratified, 
      the choice of the crossing time, while proportional to $R/\sigma$ is not straightforward. 
      It also depends on the exact efficiency of the energy injection due to the supernovae since 
      $\epsilon _{SN}$ is difficult to  estimate accurately.

         \begin{figure*}
            \begin{center}
               \includegraphics[width=17cm]{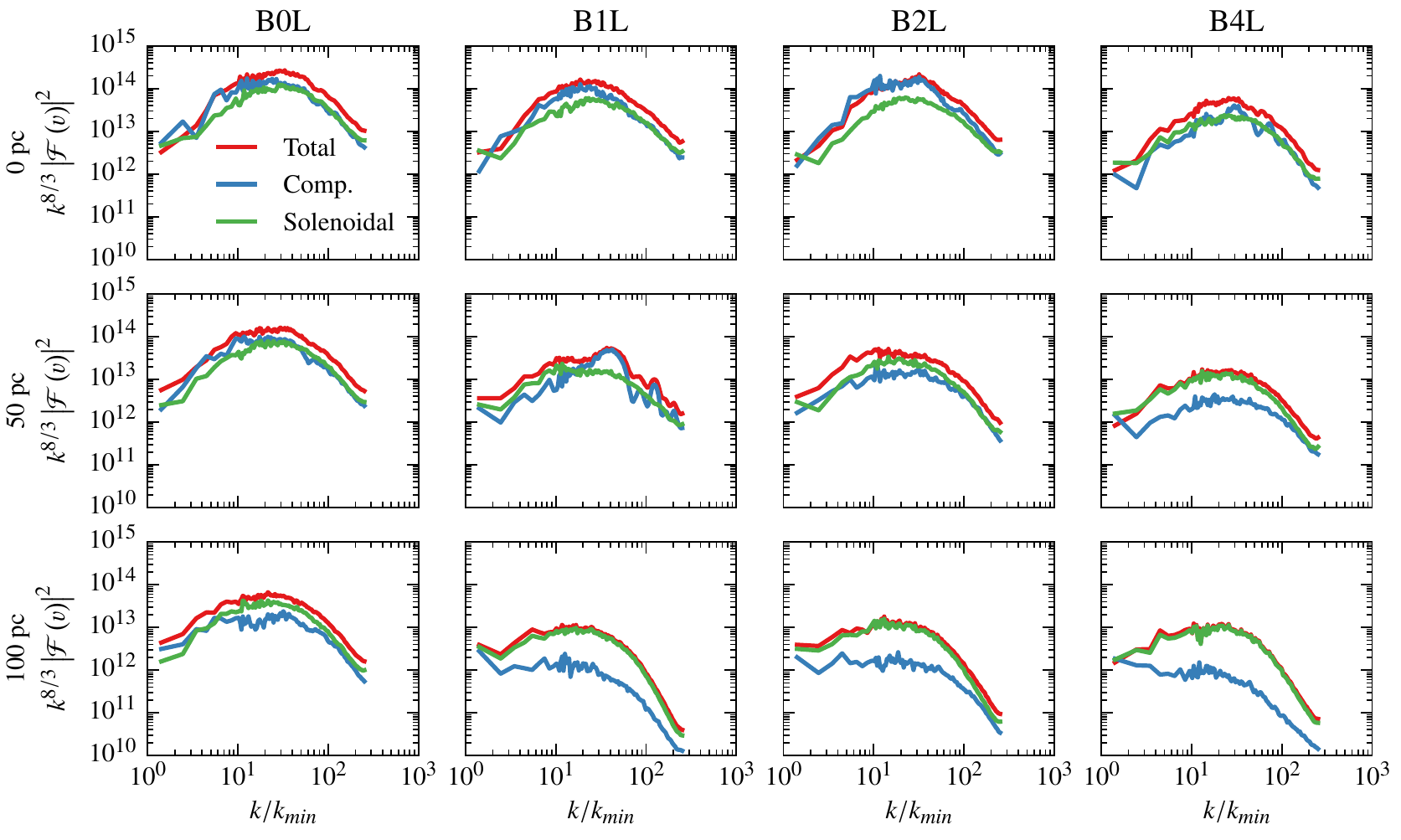}
            \end{center}
         \caption{Two-dimensional velocity power spectra with the Helmholtz
         decomposition.
         \emph{From left to right:} runs B0L, B1L, B2L, and B4L.
         \emph{From top to bottom:} altitude $0$, $50$, and $100\ \mathrm{pc}$.
         The spectra are multiplied by $k^{8/3}$ for comparison with the Kolmogorov
         scaling law.}%
         \label{fig:pspec_v_2d_s}
         \end{figure*}

      We further assume energy equipartition between
      all three directions of the velocity. Thus, we can write
      \begin{equation}
         \sigma_z^2 = \sigma^2 = \frac{1}{3} \sigma _{3D} ^2,
      \end{equation}
      where $\sigma^2$ is the total velocity dispersion in the mid-plane.
       While isotropy may not be entirely achieved given the strong stratification, 
      in the  context of protostellar formation  \citet{lee2016a,lee2016b} have investigated the differences 
      between a model which assumed isotropy and a model which considered a smaller 
      velocity dispersion along the z-axis. 
      They found that these differences remain limited. 
      %where similar energy and mechanical equilibria develop,

      Given our initial magnetic field distribution, we can assume that the
      field intensity vanishes at $z_{max}$. Then,
      \begin{equation}
         -\Delta \Pi_{mag} = \frac{\left\lvert\vec{B}_{z=0} \right\rvert^2 - 2B_{z=0,z}^2}{8\pi},
      \end{equation}
      where $\vec{B}_{z=0}$ is the magnetic field at the mid-plane and  $B_{z=0,z}$
      is its vertical component. The two terms respectively account for the
      magnetic pressure and the magnetic tension. This latter term describes the compression
      induced by the curvature of the field lines. However in our case, the field is quite 
      turbulent and it is not clear that the net effect of the $B_z$ component is captured 
      by this simple relation. As it is small anyway (Fig.~\ref{fig:mag_evol}), we will not 
      consider it further. 
      Since the magnetic field evolves in a complex way, we will simply use the values 
      displayed in Fig.~\ref{fig:mag_evol} to perform our estimate.

      To compute the gravitational energy,
      let us define the weight $\mathcal{W}$ of the gas disc with respect to a
      given gravitational potential $\phi$:
      \begin{equation}
         \mathcal{W}[\phi] = \int_0^{z_{max}} \rho(z)
            \frac{\mathrm{d}\phi}{\mathrm{d}z}(z) dz.
      \end{equation}
      We then write $\mathcal{W}_{self} = \mathcal{W}[\phi]$ and
      $\mathcal{W}_{ext} = \mathcal{W}[\phi_{ext}]$, where $\phi_{ext}$ is the
      potential given by Eq.~\eqref{eq:ext-potential}.
      Using Eq.~\eqref{eq:mhd-poisson}, we get
      \begin{equation}
         \mathcal{W}_{self} = \frac{1}{2}\pi G \Sigma^2.\label{eq:height-wself}
      \end{equation}

      In order to be able to extract the scale height of the gas disc
      analytically, we approximate the external potential $\phi_{ext}$ by
      $\phi_{ext}(z) \approx 2\pi G \rho_{ext} z^2$, where
      \begin{equation}
         4 \pi G \rho_{ext} = \left.\frac{\mathrm{d}^2 \phi_{ext}}{\mathrm{d}
            z^2}\right\rvert_{z=0} = \frac{a_1}{z_0} + a_2
      \end{equation}
      describes the density of stars and dark matter in the mid-plane, and we suppose a
      Gaussian density profile. Then,
      \begin{equation}
         \mathcal{W}_{ext} = 8 G \rho_{ext} H^2 \rho_0 = 4 G \rho_{ext} \Sigma H 
      \end{equation}

      The equilibrium equation then becomes
      \begin{eqnarray}
        \nonumber
         {1 \over 6} \left( \eta_{turb}  \dot{n}_{SN} \epsilon_{SN} E_{SN} \right)^{2/3} \Sigma^{1/3} H^{-1/3} &+&  \frac{B_0^2}{8\pi} 
         = \\
         \frac{1}{2}\pi G \Sigma^2 &+& 4G\rho_{ext} \Sigma H.
         \label{eq_model}
      \end{eqnarray}
      This equation can easily be solved using a simple bisection method, once $\eta_{turb}$, $n _{SN}$ 
      and $B_0$ are specified. 

      In order to compare the estimates of $H$ given by the above model and the
      simulations, it is useful to compute the half width at half maximum (HWHM)
      of the density distribution. Again assuming a Gaussian density profile,
      the HWHM is:
      \begin{equation}
         \Delta = \sqrt{ 2 \ln 2 } z_0 = 2 H \sqrt{\frac{\ln 2}{\pi}} \approx 0.94 \,H.
      \end{equation}

      The results are summarized and compared to the simulations in Fig.~\ref{fig:model}, 
      which shows the measured HWHM of the density profiles at $40$, $60$, and $80\ \mathrm{Myr}$ for the weak 
      and standard feedback runs (respectively red and cyan diamonds) as well as the 
      HWHM obtained by solving Eq.~(\ref{eq_model}) taking the 
      values of $\dot{n} _{SN}$ and $B_0$  from Figs.~\ref{fig:msink_all} and~\ref{fig:mag_evol}.
      Note the values of the magnetic field of the x-axis correspond to the mean magnetic field 
      in the equatorial plane at time 80 Myr for the standard feedback run (the same value is used for the 
      three snapshots and the two models for more clarity).
      The value of $\eta_{turb}$ is adjusted to match the hydrodynamical case (run B0L) at time 
      60 Myr and has been estimated to $\eta_{turb} \simeq 0.3$ (labelled model D1 in Fig.~\ref{fig:model}). 
      Note that the  results vary rather sensitively with it. 
      The corresponding models 
      show reasonable agreement except for the highest magnetization runs (B4) and at 
      time 80 Myr in the hydrodynamical and weakly magnetized cases (B0 and B1). In an 
      attempt to obtain a better fit at high magnetization, we have considered a 
      model in which $\eta_{turb}$ varies with $B$. Indeed in
      \citet{Iffrig15a} 
      we found that the momentum injected  by a
      supernova in the dense gas, does increase with $B$ by a factor on the order of 
      $\simeq 1.5$. This is likely due to the correlation that magnetic field induces
      into the flow. When a magnetized fluid particle moves it entrains 
      more fluid in its wake. 
      The model labelled D2 (dashed line in Fig.~\ref{fig:model})
      assumed $\eta_{turb} = max(0.3, B_5^2 \times 0.5)$
      where $B_5$ is the magnetic intensity in units of 5$\mu G$, that is to say the 
      most magnetized run (B4L and B4W) has $\eta_{turb} \simeq 1$. 
      As can be seen a better agreement is reached for the runs B4 at time 
      60 and 80 Myr (for run B2 the agreement is slightly better at 60 Myr and 
      worse at time 80 Myr). The disagreement at time 40 Myr persists because 
      the SFR is zero at 40 Myr for the most magnetized runs (B4). At this epoch
      the disc is probably contracting because of the flux expulsion and out 
      of equilibrium. 
      
      Finally, we see that the weak feedback cases do not agree well with the model
      at time 80 Myr. This is likely because as clearly seen in Fig.~\ref{fig:rt}
      (bottom panel), the disc is undergoing a global infall. This is also 
      consistent with the continuous increase of the SFR seen in the bottom panel of 
      Fig.~\ref{fig:msink_all} (while on the contrary the standard feedback models
      display a nearly constant SFR). 
      Therefore the simulations performed with weak feedback are out of equilibrium and the model is not expected to 
      be valid. 
      
      We conclude that a simple hydrostatic model is reasonably successful in reproducing the 
      disc thickness though not very accurate. It suggests that the supernovae driving is 
      not very efficient to inject energy in the system as the corresponding efficiency 
      required is $\eta_{turb} \times \epsilon_{SN} \simeq 0.015$. Our results 
      are compatible with  a strong field  increasing this efficiency by a factor on the order of 2-3
      though given that the overall agreement between the model and the simulation is 
      qualitative, this would need confirmation.

%         \begin{figure}
%            \begin{center}
%               \includegraphics[width=\hsize]{pspec_3d_vall_s.pdf}
%            \end{center}
%         \caption{Three-dimensional velocity power spectra in Helmholtz
%         decomposition.
%         \emph{From top to bottom:} runs B0L, B1L, B2L, and B4L.
%         The spectra are multiplied by $k^2$ (such that the Kolmogorov scaling
%         corresponds to a slope of $-5/3$)}%
%         \label{fig:pspec_v_3d_s}
%         \end{figure}

%         \begin{figure*}
%            \begin{center}
%%%               \includegraphics[width=17cm]{pspec_2d_mult_s.pdf}
%               \includegraphics[width=17cm]{2D_L_80_mult.pdf}
%            \end{center}
%         \caption{Two-dimensional power spectra for different altitudes.
%         \emph{From left to right:} runs B0L, B1L, B2L, and B4L.
%         \emph{From top to bottom:} density, magnetic field.
%         The density spectra are multiplied by $k$, and the magnetic field
%         spectra by $k^{8/3}$ (a Kolmogorov spectrum would appear flat)}%
%         \label{fig:pspec_m_2d_s}
%         \end{figure*}

      \begin{figure}
         \begin{center}
            \includegraphics[width=\hsize]{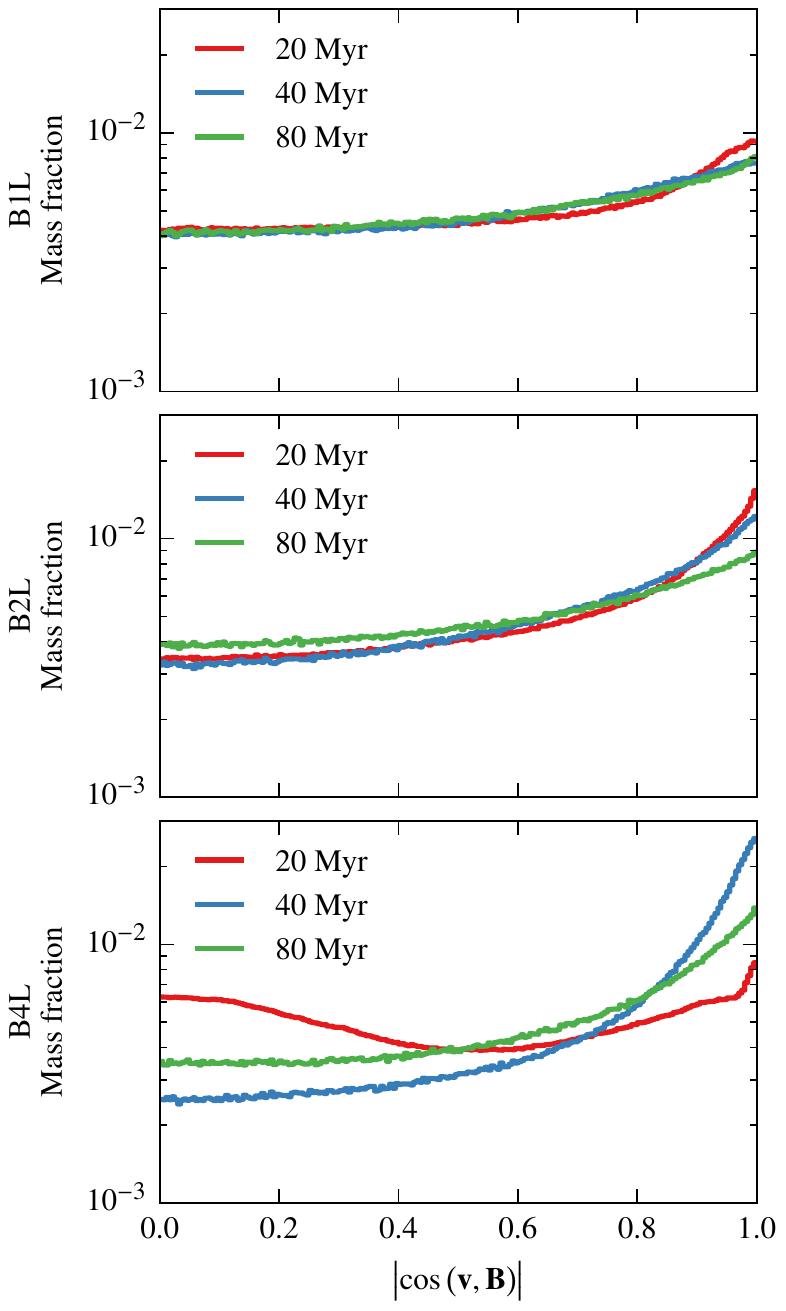}
         \end{center}
      \caption{Relative orientation of the velocity and the magnetic field as a
      function of time and magnetic field intensity for the strong feedback
      runs. The fields spontaneously align with time. The histograms are
      mass-weighted to give more importance to the gas in the galactic disc.}%
      \label{fig:v_B_angle_s}
      \end{figure}

      \begin{figure}
         \begin{center}
            \includegraphics[width=\hsize]{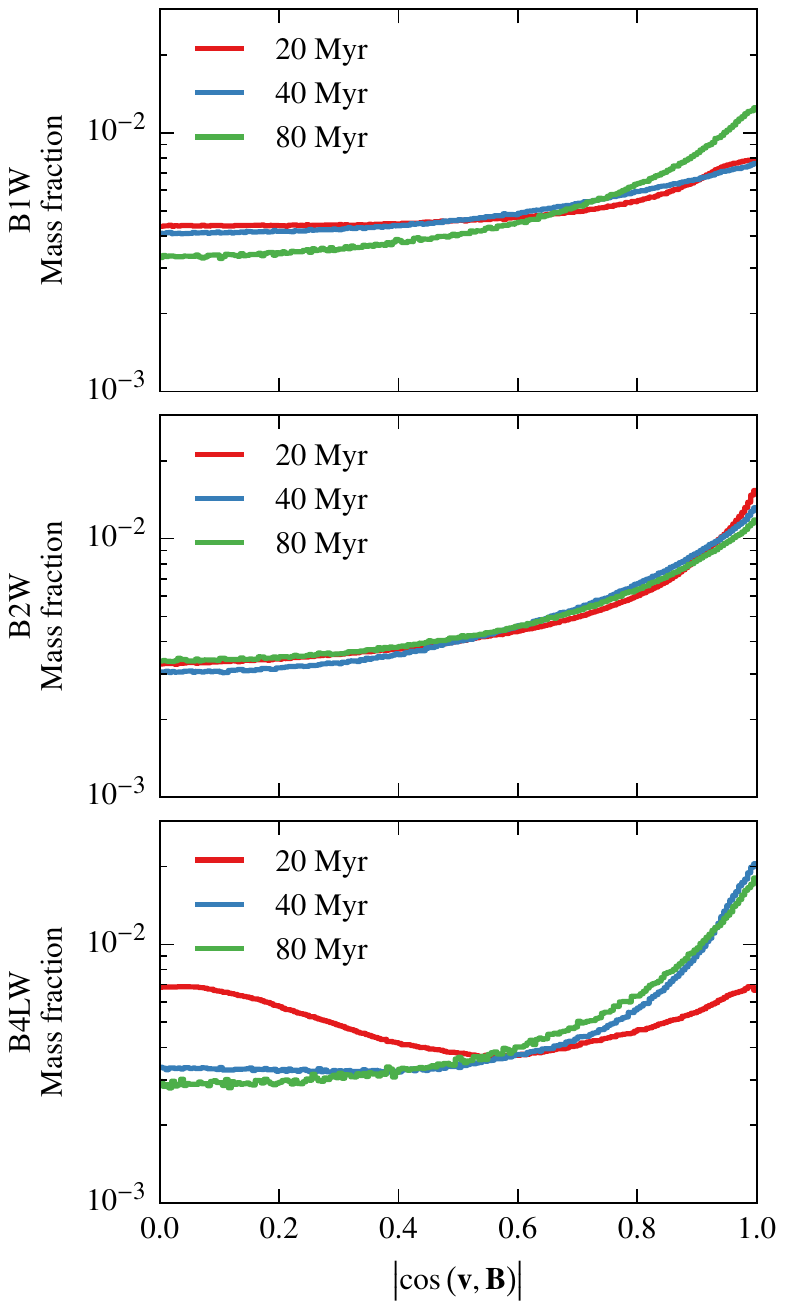}
         \end{center}
      \caption{Relative orientation of the velocity and the magnetic field as a
      function of time and magnetic field intensity for the weak feedback runs.
      The fields spontaneously align with time. The histograms are mass-weighted
      to give more importance to the gas in the galactic disc.}%
      \label{fig:v_B_angle_w}
      \end{figure}

\section{Turbulence properties}
   We now turn to the study of the detailed properties of the MHD supernova-driven turbulence 
   which takes place in the stratified galactic disc. 

%__________________________________________________________________

   \subsection{Density distribution}

   We first consider the density distribution as it is a fundamental parameter 
   for the ISM physics and the star formation process. 
      Previous studies found that  the density distribution
      associated to supersonic isothermal turbulence  is log-normal \citep{Vazquez94,
      Kritsuk07, Federrath08, Federrath10}, with a power-law tail due to gravity
      \citep{Kritsuk11}. 
      Larger-scale simulations \citep[e.g.][]{Hill12, Kim13, Hennebelle14} find
      such a log-normal distribution only for the gas with density typically above $10\
      \mathrm{cm^{-3}}$, which corresponds to the cold phase 
      (the densest gas showing the power-law
      tail is not visible due to the lack of resolution).

      Top panel of Fig.~\ref{fig:n_hist} displays the density PDF at time 40 Myr for the 
      runs B0, B1 and B4. It shows a good agreement with a log-normal
      distribution for densities above $10\ \mathrm{cm^{-3}}$. Given the
      presence of sink particles, densities above $1000\ \mathrm{cm^{-3}}$ are
      not present in the simulations. The most diffuse gas is a consequence of the
      supernova feedback, and the dip around $1\ \mathrm{cm^{-3}}$
      is due to the thermal  instability. 
      %When the magnetic field strength increases, the density
      %distribution becomes narrower \citep[see also][]{Molina12}, resulting in a
      %reduced amount of dense gas, hence the lower star formation rate.

   %__________________________________________________________________

       Since column densities are observationally inferred, though usually 
         for individual clouds \citep[e.g.][]{kainulainen2009,schneider2015}
        rather than for a fraction of the galactic plane, we also 
      present them here for future references and comparisons. 
      Bottom panels show the column density distributions obtained 
      by integrating face-on and edge-on through the galactic plane. The shape 
      is also broadly lognormal. As can be seen the most magnetized runs 
      present the highest values, which is a consequence of the dense 
      structures being larger.

   \subsection{Power spectra}

      In order to understand the details of the supernova-driven turbulence in
      the simulations, we take advantage of the uniform resolution to compute
      power spectra. Since the simulations include a stratified structure, we
      calculate both the full three-dimensional spectra and two-dimensional
      spectra on slices of constant altitude.

   %__________________________________________________________________

%      \subsubsection{Method}
%         We estimate the power spectrum of various quantities by computing the
%         discrete Fourier transform of the whole data cube:
%         \begin{equation}
%         \hat{q}_{lmn} = \sum_{x,y,z} q_{xyz} \exp\left( -2 i \pi \frac{lx + my + nz}{N} \right).
%         \end{equation}
%         The power is defined as the square magnitude of the Fourier transform. We
%         then average between spheres of constant wave vector magnitude to get a
%         one-dimensional power spectrum.
%         \begin{equation}
%         P_q(k, k') = \frac{1}{\mathcal{N}(k, k')} \sum_{k \le \sqrt{l^2 +
%            m^2 + n^2} \le k'} \left|\hat{q}_{lmn}\right|^2,
%         \end{equation}
%         where $\mathcal{N}(k, k')$ is the number of wave vectors with a magnitude
%         between $k$ and $k'$.

         For the velocity, we also compute a Helmholtz decomposition between
         compressive (vanishing curl) and solenoidal (vanishing divergence) modes
         by projecting the Fourier transform of the velocity parallel and
         perpendicular to the wave vector: $\vec{v} = \vec{v}_{comp} +
         \vec{v}_{sol}$ where $\hat{\vec{v}}_{comp}(\vec{k}) \parallel \vec{k}$ and
         $\hat{\vec{v}}_{sol}(\vec{k}) \perp \vec{k}$. For the 2-dimensional power
         spectra, these quantities are computed on the whole 3D Fourier cube, and
         then an inverse Fourier transform is performed along the vertical axis. 
         Then the power spectrum of the corresponding field is computed at a given
         altitude. 

         While power spectra of turbulent flows in the context of the ISM have been widely 
         explored (see references below), it is important to study them in our simulations
         i) for a consistency check, ii) because there are specific aspects such as the stratification
         that have not been widely explored. 

   %__________________________________________________________________

      \subsubsection{Three-dimensional power spectra}

         Contrary to most studies of turbulence where the energy is injected at
         large scales, for instance by an external force at wave vectors $1 \le
         k < 3$ \citep[e.g.][]{Kritsuk07, Federrath12,federrath2013},  the turbulence in 
         our simulation is 
         supernova-driven, and  energy is likely injected at
         scales on the order of 50 pc although since supernovae
         are correlated, they may also inject energy at larger scales. 
         This work \citep[and this paper]{Iffrig15b} 
         present similarities with the
         work by \citet{Padoan16}, but differs from it 
         for the following points:
           i) their $250\ \mathrm{pc}$ box does not include vertical
            stratification and is periodic, ii)
           the supernovae are not correlated to star-forming regions,
            iii) the resolution is not uniform ($128^3$ base grid).
         This will introduce some discrepancies that will be discussed. Given
         our resolution, the wave vectors $k$ above $64 k_{min}$ (where $k_{min}
         = 2\pi / L$, $L = 1\ \mathrm{kpc}$ being the size of the simulation
         box) are affected by the numerical dissipation. Besides, the
         large-scale stratification will affect the three-dimensional power spectra 
         at low $k$. Therefore, the power-law scalings are accurate only
         for (roughly) $10 < k / k_{min} < 64$.
         The three-dimensional power spectra for various variables are shown on
         Fig.~\ref{fig:pspec_m_3d_s}. Since the most magnetized runs have been 
         performed at a resolution of 512$^3$, we show here the runs B0L-B4L. 
         For reference the high resolution simulation power spectra are given in appendix~\ref{appen-3D}.

         The
         velocity power spectra show reasonable agreement with the well-known
         models \citep{Kolmogorov41, Iroshnikov64, Kraichnan65, Sridhar94,Lee2010,grappin2010,mason2012,beresbyak2011},
         without allowing to discriminate between the two slopes $k^{-7/2}$
         and $k^{-11/3}$. In the context of supersonic turbulence 
         power spectra on the order of $k^{-3.9}$ have been inferred \citep[e.g.][]{Kritsuk07},
         which has been interpreted as a consequence of compressibility. In the 
         present case, the somewhat shallower power spectrum may be interpreted as a consequence 
         of the fact that most of the gas is warm and has Mach number on the order of 1
         meaning that the effect of compressibility 
         could be less pronounced that in high Mach number flows. 

         The slopes of the density power spectra are very flat and compatible in the inertial 
         range with $E(k) \propto k^0$, which is usually interpreted as a consequence 
         of the steep variations due to supersonic shocks and thermal instability since 
         the Dirac function has a power spectrum with an index equal to 0.
         Interestingly, the slope of the hydrodynamical runs (B0L) is slightly shallower
         than the MHD ones. The power spectra of $\log \rho$ are much  shallower
         and present an index and a general behaviour close to the 
         velocity power spectra \citep{schmidt2009,Audit2010}. 

%         in the magnetized runs are in remarkable agreement with the model given
%         by \citep{Fleck96} with $\alpha = 1/6$, which corresponds to a fractal
%         dimension of $D = 2.5$. The run without magnetic field B0L shows a slope
%         that corresponds to $\alpha = 0.26$ and $D = 2.23$, which can be
%         interpreted as a slightly more isotropic compression of the gas. 

         The density-weighted velocity $\rho^{1/3} v$, supposed to be the
         compressible equivalent of the velocity  for the power spectra 
         \citep[see][]{Kritsuk07}, has a slope that is even shallower than 
         the velocity field power spectrum. At small $k$ its value approaches 
         $k^{-2}$. At intermediate scales it becomes stiffer and approaches
         a slope on the order of $k^{-11/3}$ between $k=10$ and 100, 
         particularly for the magnetized runs, although the limited 
         range does not allow a solid conclusion. This behaviour is a possible 
         signature of the injection of turbulence at scale of about 100 pc. 
         Another alternative explanation is that this signs the transition from 3D turbulence 
         to a more 2D one since the scale height of the disc is about one tenth of the total 
         box length. In this latter case, an energy power spectrum $k^{-8/3}$ would be expected.
         Let us stress that the enstrophy cascade that would lead to $k {\cal F}^2 \propto k^{-3}$ 
         requires conservation 
         of the vorticity which is not ensured in the magnetized and non-barotropic flows like the ISM
         \citep{Hennebelle2007,Padoan16}.

         Finally we note that in the magnetized simulations, the magnetic field power spectra
         show the same behaviour as the velocity field ones, a behaviour found 
         in turbulent periodic boxes \citep{kritsuk2011}.

%__________________________________________________________________

      \subsubsection{Two-dimensional power spectra}

         The vertical stratification of our simulations induces strong contrasts
         as a function of altitude. Therefore, we computed two-dimensional power spectra 
         on several horizontal planes to get a more detailed view of
         the consequences of this stratification. For most quantities like 
         the density and the magnetic field, the power spectra do not look too different 
         at different altitudes apart from their amplitude. 
         Therefore for the sake of conciseness,
         we focus on the velocity power spectra, shown on
         Fig.~\ref{fig:pspec_v_2d_s}. The two-dimensional power spectra of 
         other quantities are shown in appendix~\ref{appen-2D}.

         Star formation simulations with turbulent forcing \citep{Federrath12}
         have shown that the star formation rate can be reduced by an order of
         magnitude between a purely compressive and purely solenoidal stirring.
         Energy equipartition between those two components would show up as a
         ratio $P_{sol} / P_{comp} \approx 2$, since there are two solenoidal
         modes for one compressive. 

         In the galactic mid-plane, Fig.~\ref{fig:pspec_v_2d_s} shows that the
         compressive and solenoidal power spectra are comparable, 
         which means that the compressive modes are a factor of
         about $2$ higher than expected if energy equipartition was achieved. 
         According to the results of \citet{Federrath12}, this means
         that star formation is neither in the most favorable regime (which
         would be purely compressive), nor in the least favorable (purely
         solenoidal). 
         There is a possible trend that in the galactic plane
         at $z=0$ the compressible mode power spectra are slightly steeper than the solenoidal one
         although the lack of statistics makes  the  noise level significant.
         This is broadly compatible with the work of \citet{Padoan16}
         in which $P_{sol} \propto k^{-3.31}$ and $P_{comp} \propto k^{-3.98}$. 
         For the compressive modes, this corresponds to a
         Burgers (pressureless) fluid. 

         At higher altitudes, the solenoidal modes dominate, and the ratio of
         solenoidal to compressive power is stronger for higher magnetic field.
         This is expected since the magnetic field helps creating solenoidal motions
         and tends to impede the gas compression.

         These results are at variance   with the work of
         \citet{Padoan16} since they find that the solenoidal 
         modes dominate while we find that this depends on altitude 
         but in the equatorial plane the compressible modes are dominating.         
         One major difference with this work is the stratification induced by the 
         galactic gravitational field. Another important one is the 
         supernova scheme: they use a random supernova distribution, similar to
         scheme A of \citet{Hennebelle14}, which injects supernovae mostly in
         the diffuse gas. 
         %Therefore, it is not surprising to find such a steep
         %compressive power spectrum, since the pressure in the diffuse gas is
         %(at least near thermal equilibrium) much lower than that of the dense
         %gas. 
         The solenoidal power spectrum is slightly shallower than the
         Kolmogorov power spectrum, which is comparable to our results.
         It must be kept in mind that the exact correlation between supernova explosions
         and the dense gas is not well constrained. There is a  possibility that the scheme 
         used here is not accurate enough and that the supernovae should be less 
         correlated with the dense gas. However in this case, the SFR is far too high
         as stressed in \citet{Hennebelle14}.
         Generally speaking, the exact way feedback 
         operates remains to be clarified. One very important constraint
         is that the SFR should be compatible with the observational rates.

%__________________________________________________________________

   \subsection{Alignment of velocity and magnetic fields}
      
      Star formation simulations based on colliding flows \citep{Inoue09,Inoue2012,kortgen2015} 
      show the importance of the alignment between velocity and
      magnetic field: if the inflow velocity is not along the magnetic field
      lines, star formation is reduced efficiently. This is expected since the 
      transverse component of the magnetic field is amplified by the converging
      velocity field. However, it is known that in magnetized flows 
      the velocity and magnetic fields tend to align  \citep{Boldyrev06, Matthaeus08,Banerjee09}.
      This effect is due in part to the Lorentz force which vanishes along the magnetic field 
      and therefore is expected to be smaller when the magnetic and velocity fields are parallel. 
      It is also a consequence of how the velocity and magnetic field get transported 
      and generated by the flow. 

      To what extent  colliding flow calculations with an inclined magnetic field 
      are representative of the ISM remains to be clarified. To understand the exact role 
      magnetic field is playing in the ISM, knowing its orientation  with respect 
      to the velocity field is crucial. 
      For this purpose, we study the angle between
      velocity and magnetic fields, defined as:
      \begin{equation}
         \cos \left(\vec{v}, \vec{B}\right)
            = \frac{\vec{v} \cdot \vec{B}}%
                   {\left\lvert\vec{v}\right\rvert
                   \left\lvert\vec{B}\right\rvert}.
      \end{equation}
      A uniform distribution of relative orientations would lead to a uniform
      distribution of this cosine.

       The results are shown on
      Figs.~\ref{fig:v_B_angle_s} and~\ref{fig:v_B_angle_w} where we see that the 
      mass weighted angle distribution  clearly shows an excess in the aligned 
      configuration. The amplitude of the effect  increases with the magnetic 
      intensity and the feedback strength. For the B1 run, the distribution is 
      2 times higher for $\cos \left(\vec{v}, \vec{B}\right)=1$ than 
      $\cos \left(\vec{v}, \vec{B}\right)=0$, this value is about 3-4 for run B2 
      and 6-10 for run B4. For runs B1W, B2W and B4W, the effect is even more pronounced
      which demonstrates that feedback is playing an important role there. Stronger 
      feedback tends to dealign the velocity and magnetic fields which is expected since 
      in a super-Alfv\'enic shock, the transverse component of the magnetic field is 
      amplified and the velocity tends to be perpendicular to it. Interestingly, we 
      see in runs B1, B2 and B4 that the distribution at time 80 Myr shows 
      less alignment that at time 40 Myr, which is also consistent with feedback
      reducing the alignment.  This effect was reported in \citet[]{Passot95}.
     
       To get a more accurate understanding of the magnetic and velocity alignment 
      we have also  investigated the dependence of the alignment with 
      gas density in appendix~\ref{BV_rho}. 

      Altogether these results show that the magnetic field and the velocity field 
      are well correlated and that it is necessary when estimating the 
      magnetic field influence to consider this effect. In particular, 
      even in the case of the less magnetized runs (B1) about half of the 
      converging flows are expected to present an angle that is below 45$^\circ$.
      This number is even higher for the more magnetized runs.

%      \begin{figure*}
%         \begin{center}
%            \includegraphics[width=17cm]{vB_z_s.pdf}
%         \end{center}
%      \caption{Relative orientation of the velocity and the magnetic field as a
%      function of time, altitude and magnetic field intensity for the strong
%      feedback runs. The fields spontaneously align with time. The histograms
%      are mass-weighted.
%      \emph{From left to right:} runs B1L, B2L and B4L (increasing initial magnetic
%      field).
%      \emph{From top to bottom:} altitudes $\left|z\right| < 50\ \mathrm{pc}$,
%      $50\ \mathrm{pc} \le \left|z\right| < 100\ \mathrm{pc}$,
%      $100\ \mathrm{pc} \le \left|z\right| < 200\ \mathrm{pc}$,
%      $\left|z\right| \ge 200\ \mathrm{pc}$.
%      }%
%      \label{fig:v_B_z_s}
%      \end{figure*}

   \begin{figure*}
      \begin{center}
         \includegraphics[width=17cm]{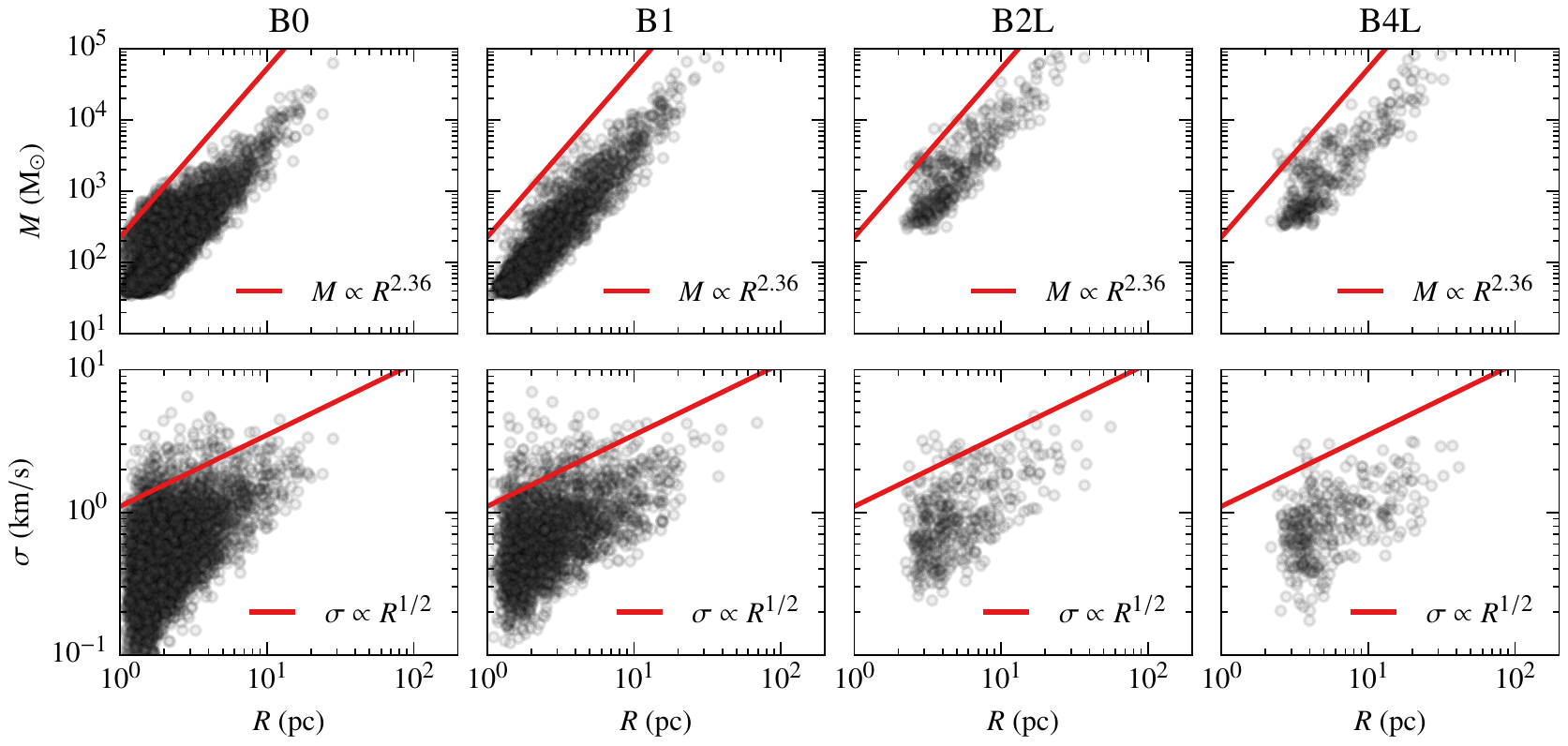} \\[1ex]
         \includegraphics[width=17cm]{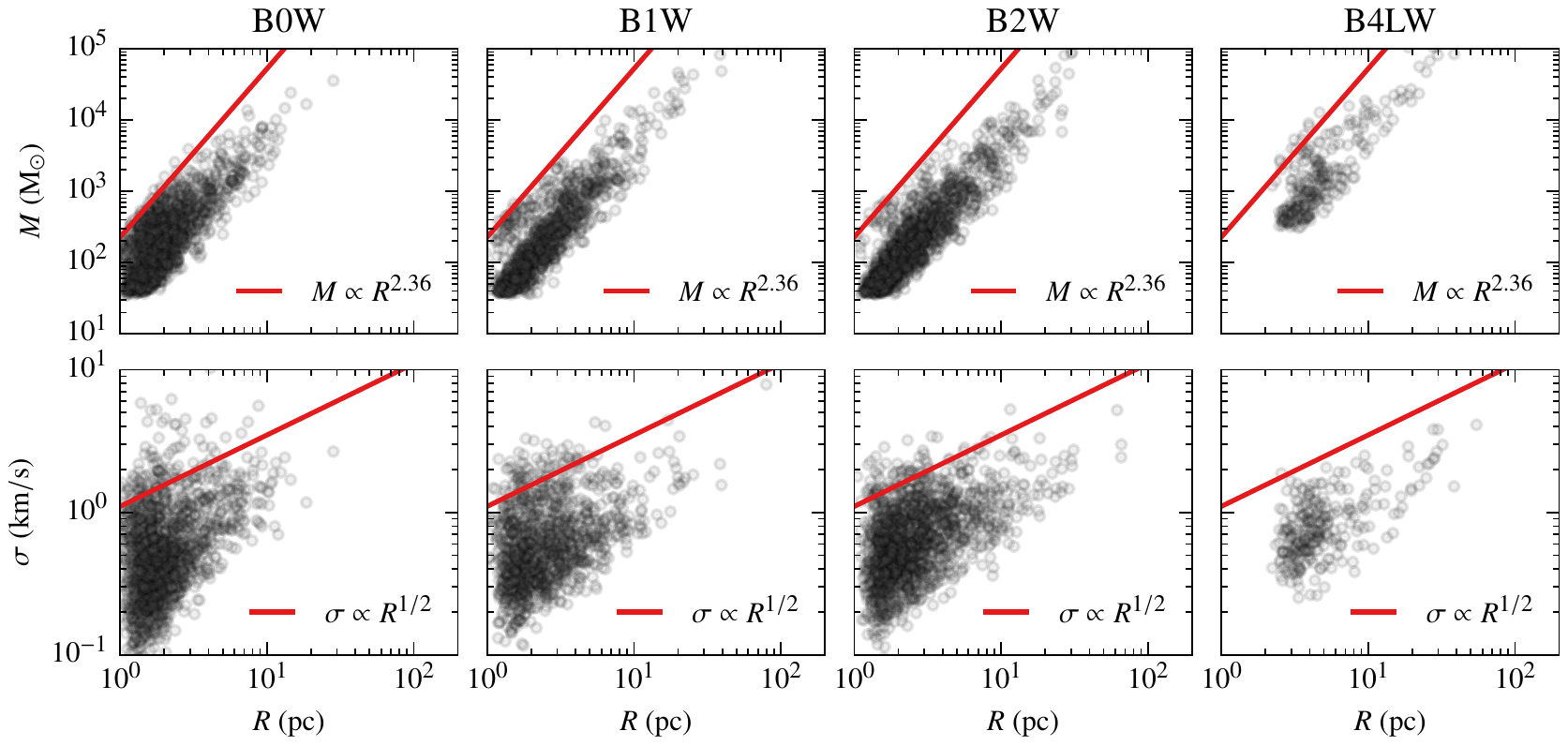}
      \end{center}
   \caption{Clump scaling relations at $60\ \mathrm{Myr}$.
   \emph{From left to right:} runs B0, B1, B2, and B4.
   \emph{First row:} mass-size relation. \emph{Second row:} size-velocity
   dispersion relation.
   \emph{Top panels:} strong feedback. \emph{Bottom panels:} weak feedback.
   The solid red lines show the power-laws stated by Eq.~(\ref{powerlaw}).
   }\label{fig:larson}
   \end{figure*}

   \begin{figure*}
      \begin{center}
         \includegraphics[width=17cm]{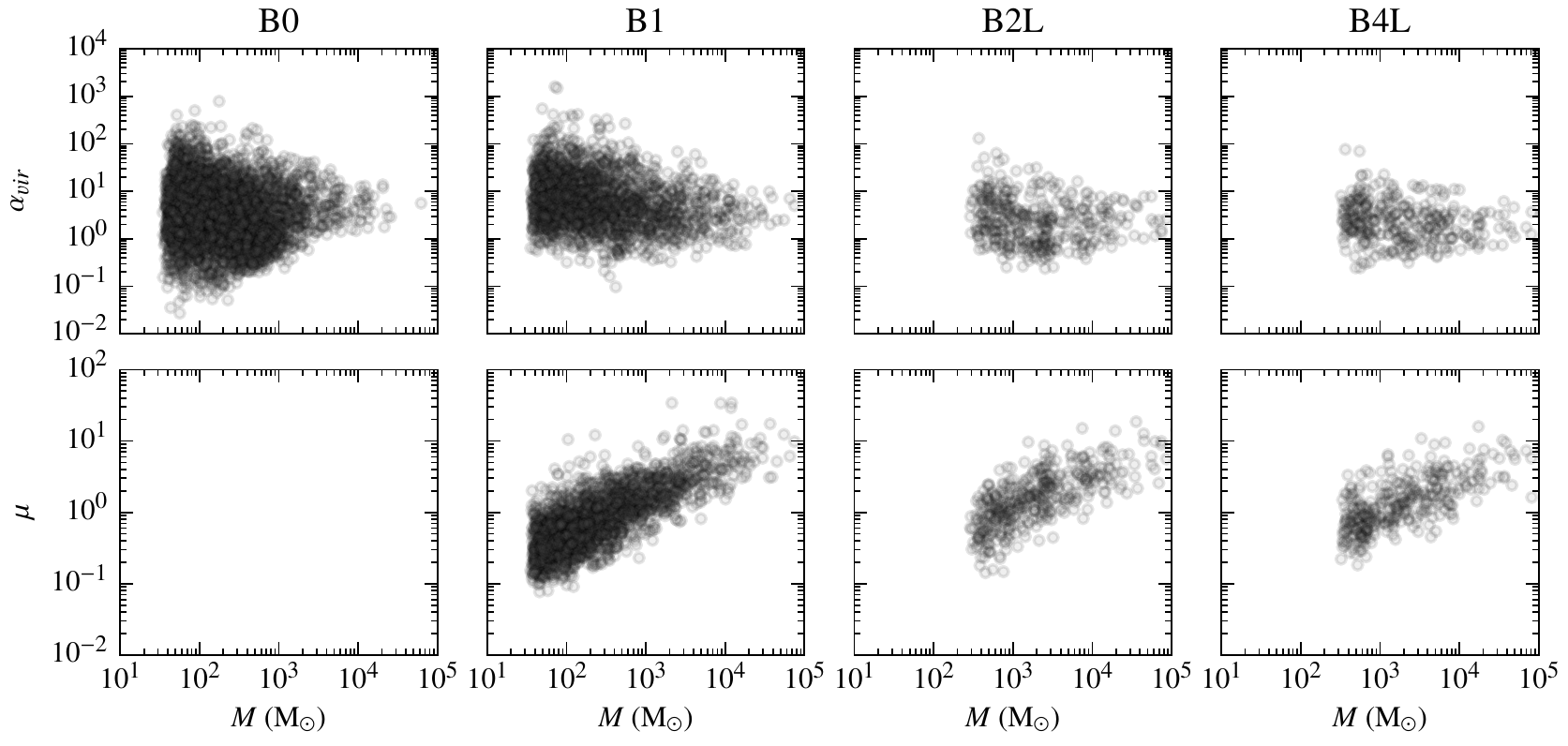} \\[1ex]
         \includegraphics[width=17cm]{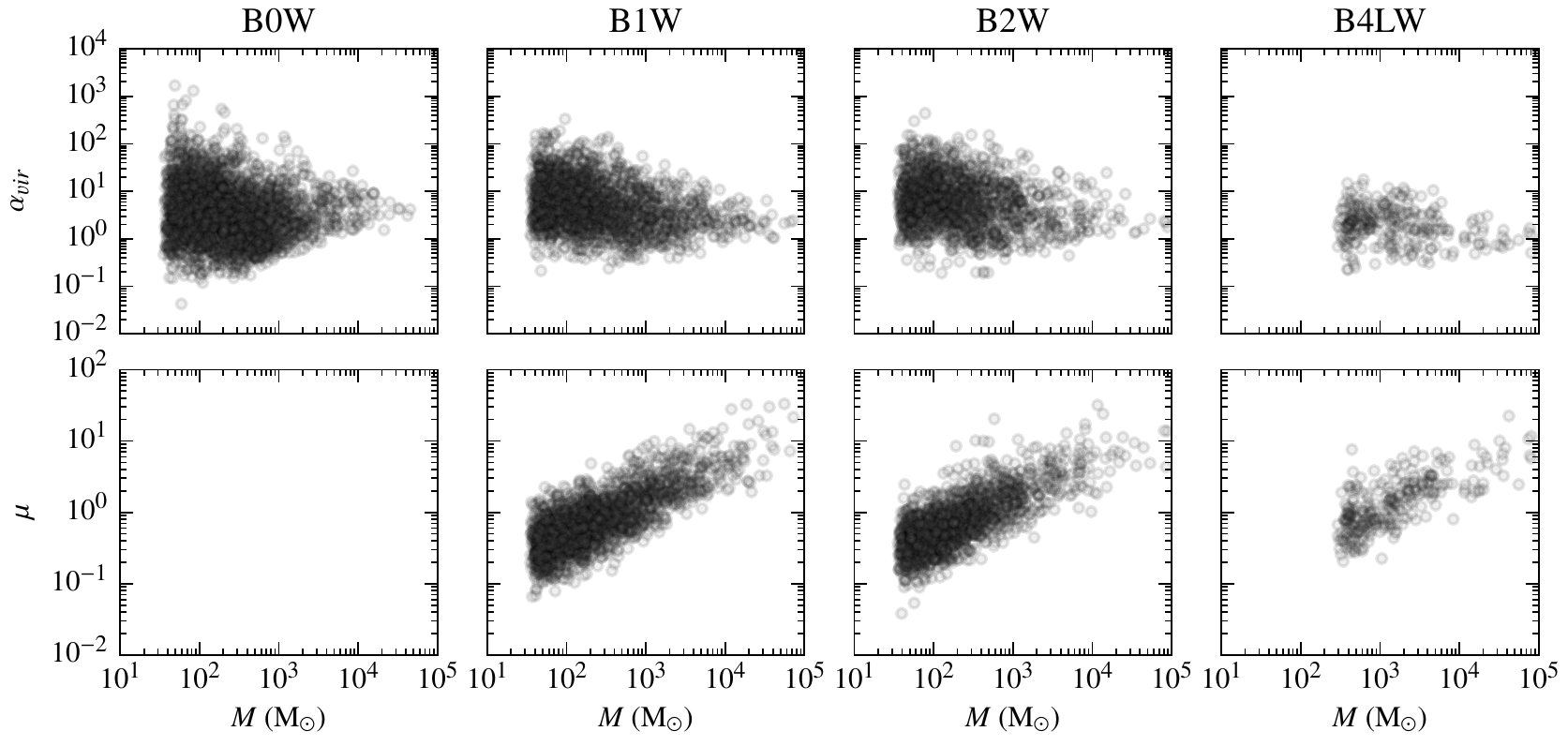}
      \end{center}
   \caption{Virial $\alpha$ parameter and mass-to-flux over critical 
     mass-to-flux ratio of clumps  at $60\ \mathrm{Myr}$.
   \emph{From left to right:} runs B0, B1, B2, and B4.
   \emph{First row:} mass-$\alpha$ relation. \emph{Second row:} mass-$\mu$
    relation.
   \emph{Top panels:} strong feedback. \emph{Bottom panels:} weak feedback.
   }\label{fig:alpha-mu}
   \end{figure*}

\setlength{\unitlength}{1cm}
   \begin{figure*}
      \begin{picture}(0,15)
         \put(0,7){\includegraphics[width=10cm]{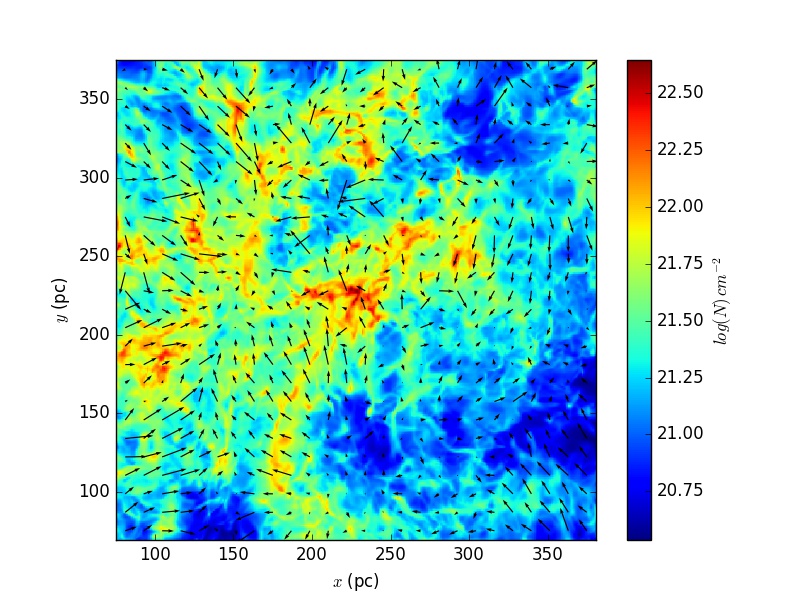}}
         \put(9.5,7){\includegraphics[width=10cm]{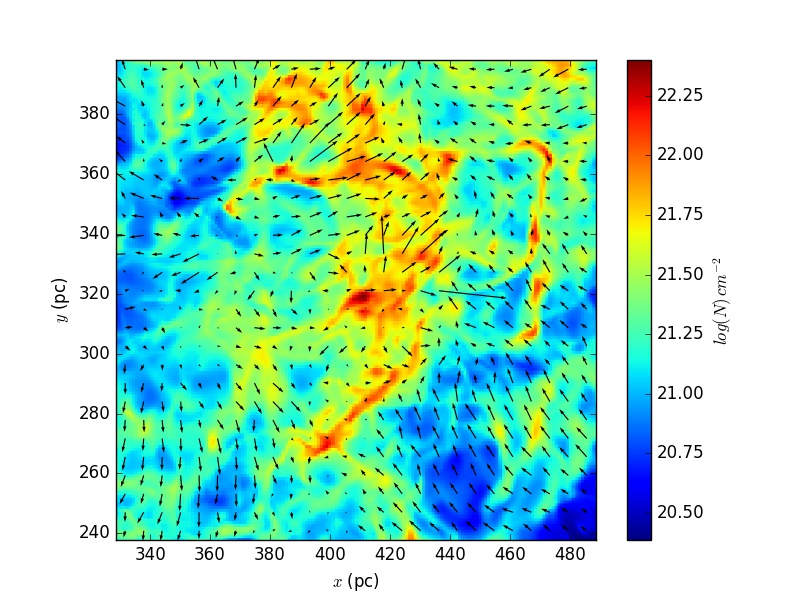}} 
         \put(0,0){\includegraphics[width=10cm]{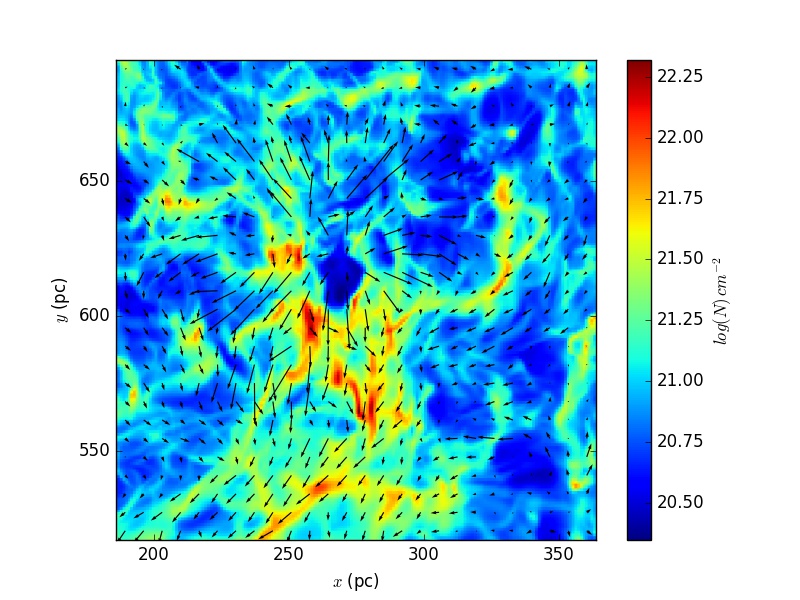}} 
         \put(9.5,0){\includegraphics[width=10cm]{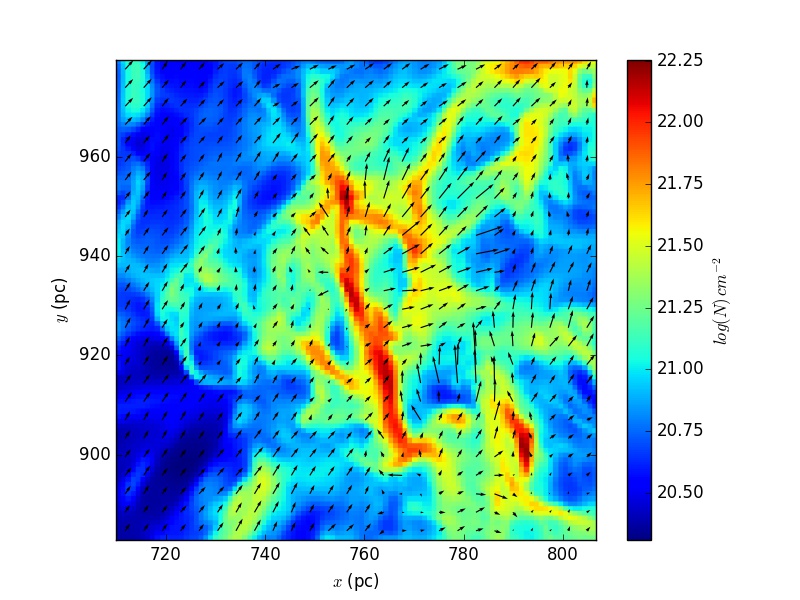}} 
      \end{picture}
   \caption{Images of the four most massive clouds identified in the simulation B1 at time 60 Myr. 
     Column density and mean velocity field in the z-plane. 
   }\label{clump}
   \end{figure*}

%__________________________________________________________________

\section{Structure formation}
   We now turn to the dense clouds, which form under the influence of gravity and turbulence. 
   We are primarily interested here by getting a statistical description and to 
   determine to what extent their properties vary with feedback and magnetization. 
   In particular the existence of scaling relations, the so-called Larson relations, is well established
    \citep[][etc.]{Larson81, Solomon87, Falgarone09, heyer2009, Roman10, Hennebelle12, miville2016}:
   \begin{align}
      \sigma &= \sigma_0 \left( \frac{L}{1\ \mathrm{pc}} \right)^\alpha,\\
      M &= M_0 \left( \frac{L}{1\ \mathrm{pc}} \right)^\beta,
      \label{powerlaw}
   \end{align}
   where $L$, $\sigma$ and $M$ are respectively the size, velocity dispersion
   and mass of the clumps. Typical values of these parameters are $\sigma_0 =
   1.1\ \mathrm{km / s}$, $\alpha = 0.5$ \citep{Falgarone09}, $M_0 = \left(228
   \pm 18\right) \mathrm{M_\sun}$ and $\beta = 2.36 \pm 0.04$ \citep{Roman10}.
   Note that these observations are performed in the CO lines, which 
   is typically tracing molecular gas of densities on the order of few 10$^2$ cm$^{-3}$.
      \citet{heyer2009,miville2016} infer a relation that entails the cloud 
   column density, namely:
   \begin{eqnarray}
     \sigma = 0.23 \; {\rm km \, s^{-1}} (\Sigma R)^{0.43}. 
   \end{eqnarray}
   where $\Sigma R$ is expressed in $M_\odot$ per pc.
   They show that this distribution present less dispersion 
   that the $\sigma-R$ one and even less than the $\sigma-M$ ones, suggesting 
   that gravity could be playing an important role here. We note that the effect 
   is clear for massive structures but less apparent for low mass ones.

   The structures are obtained with a density threshold
   of $50\ \mathrm{cm^{-3}}$ using a simple friends of friends algorithm.
   The reason for this threshold is that at a density  of  10$^3$ cm$^{-3}$ 
   the sink particles are being introduced. Therefore, to get a significant dynamical 
   range, we adopt a value that is well below. This means that we may not be tracing 
   exactly the same gas. However,  observations of the atomic gas have also been performed
   in external galaxies as the LMC \citep{kim+2007} and revealed similar behaviour (though 
   different numbers are obviously inferred).  In principle structures should be identified 
     in the same way that observers proceed. However this would imply several steps and in particular 
     the calculation of the CO molecules abundances \citep[e.g.][]{duarte2015}. This latter point is particularly difficult because 
   the CO abundances predicted by PDR codes for intermediate density gas (column densities smaller than a few
    10$^{20}$ cm$^{-2}$) are underestimated
   by almost one order of magnitude (see Fig.~11 of \citet{levrier2012}). These issues would require a dedicated study and are clearly beyond the 
   scope of the paper.

   We define the velocity dispersion, $\sigma$, the cloud radius, the virial $\alpha$ parameter 
   and the mass-to-flux over critical mass to flux ratio \citep{mouschovias1976} as 

   \begin{eqnarray}
     \nonumber
     {\bf v_0} = {\sum {\bf v} \rho dx \over \sum \rho dx }, \\
     \nonumber
     \sigma ^2 = {1 \over 3} {\sum ({\bf v} - {\bf v_0})^2 \rho dx \over \sum \rho dx } , \\
     \label{def_stat}
     R = \left( {\lambda _1 \lambda _2 \lambda _3 M^{-3}  } \right)^{1/6}, \\
     \nonumber
     \alpha = {5 \sigma^2 R \over G M}, \\
     \nonumber
     \Phi = \sum B dx^2, \\
     \nonumber
     \mu = {M  \sqrt{G} \over 0.13 \Phi}.
   \end{eqnarray}
   To compute $\Phi$, the magnetic flux, we first compute the cloud center of mass, then we compute the flux accross the 
   three planes parallel to $xy$, $xz$ and $yz$ and passing through the center of mass. We then take the largest of these 
   three fluxes.  To compute the radius, $R$, we use the eigenvalues, $\lambda_i$, of the inertia matrix, $I_{ij}$ defined by
     \begin{eqnarray}
       \nonumber
       I_{11} = \sum (y^2+z^2)dm, I_{22} = \sum (x^2+z^2)dm, \\ 
       I_{33} = \sum (x^2+y^2)dm,
       I_{12} = I_{21} = -\sum xy dm,  \\
       I_{13} = I_{31} = -\sum xz dm, I_{32} = I_{23} = -\sum yz dm.
       \nonumber
    \end{eqnarray}
     
     While this choice is reasonable, it is not unique and we have tried different definitions such as 
     using the largest eigenvalues of the mean spherical radius $R = (V/(4 \pi /3))^{1/3}$ and the resulting 
     distributions do not vary very significantly. 

   \begin{figure*}
      \begin{center}
         \includegraphics[width=17cm]{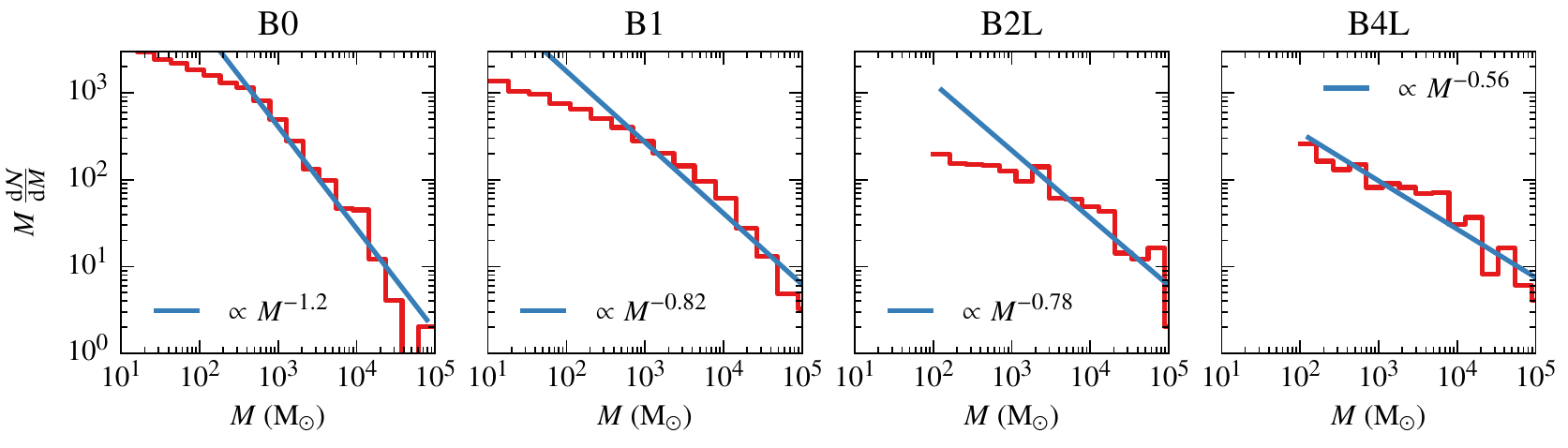}\\[1ex]
         \includegraphics[width=17cm]{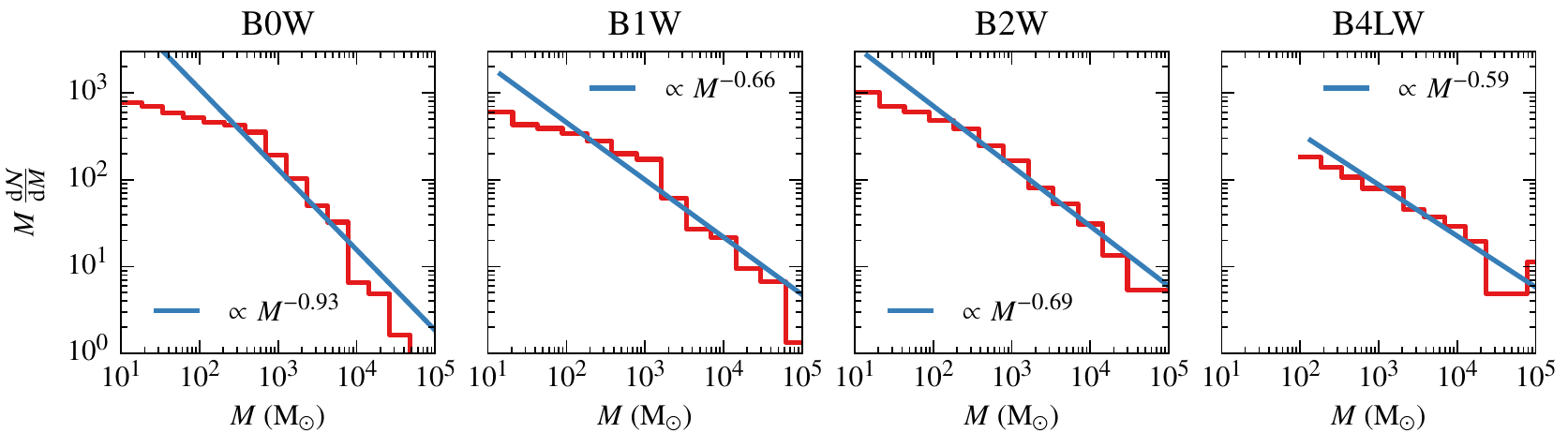}
      \end{center}
   \caption{Clump mass spectra at $60\ \mathrm{Myr}$. The lines represent the
   best power-law fit for $M > 100\ \mathrm{M_\odot}$.
   \emph{From left to right:} runs B0, B1, B2, and B4.
   \emph{First row:} strong feedback. \emph{Second row:} weak feedback.
   }\label{fig:mclump}
   \end{figure*}

   The mass-size and size-velocity dispersion relations are shown on
   Fig.~\ref{fig:larson}. 
   The solid red lines show the power-laws stated by Eq.~(\ref{powerlaw}).
    As can be seen the mass-size relation of the structures follow a similar power-law to the 
   observed one with a slope on the order of 2.3. 
   The total mass is below the  one inferred from CO survey but as 
   explained above it is likely a consequence of the density threshold being too low. To 
   verify this, we have extracted the clumps using a threshold of 200 cm$^{-3}$. In this 
   case the cloud masses is as expected about 4 times larger and present the same 
   power-law behaviour. 
   Interestingly, the number of small clumps is much higher in the hydrodynamical run B0
   than in the MHD ones and decreases with magnetic intensity, while the power-law 
   behaviour does not change. This effect, which has already been noticed in 
   smaller-scale simulations \citep{hennebelle2013} is likely a consequence of magnetic tension,
   which makes the flow more coherent. 
     The velocity dispersion is also displayed in  Fig.~\ref{fig:larson}. The values present 
   a significant dispersion. The largest velocity dispersion of the  clouds in simulations B0, B1 and B2 
    are comparable with the largest velocity dispersion inferred from observations \citep{Falgarone09}.
    Both distributions present a large spread and some clouds have a velocity dispersion 
    significantly below the mean value. For the sake of completeness, the velocity dispersion, $\sigma$ as a function of $R \Sigma$ is shown in 
    appendix~\ref{appen-sig-Rsig}.
    As can be seen the velocity dispersion is significantly  lower in the most magnetized case, simulation B4. 
    This is a consequence of stronger field which makes it difficult to bend the field lines but also 
    likely of the reduced star formation rate (as shown in Fig.~\ref{fig:msink_all}).  Interestingly the 
    simulations with weak feedback B0W-B4W present very similar properties to the standard feedback case. 
    This is indeed expected since, as discussed before, the amount of momentum delivered in the ISM are comparable 
    because the SFR are higher in the weak feedback simulations. 
    
    Figure~\ref{fig:alpha-mu} displays the virial $\alpha$ parameter, which allows to quantify the importance
    of gravity in the cloud. The run B0 (hydrodynamical and standard feedback) presents a broad distribution 
    of nearly three decades for 100 M$_\odot$ clouds for which $\alpha$ goes from 100 to 0.1, though typical 
    values are on the order of 10. The distribution is narrower for larger masses and for clouds 
    of mass larger than $10^4$ M$_\odot$, it typically ranges from about 1 to 10 (note that exact value depends on the 
    chosen definition of the radius, $R$). The run B1 shows similar trends except that the spread is significantly reduced 
    for clouds of small masses. The same is true for runs B2L and B4L except that since they have been performed at 
    a lower resolution, 100 M$_\odot$ clouds are absent. The behaviour  for the weak feedback runs is also similar 
    with a trend for slightly lower values.  Altogether these results suggest that the turbulence within 
    molecular clouds is not primarily due to gravity for most low-mass clouds simply because 
    the values of $\alpha$ can be larger than 10 and are on average larger than for the most massive clouds. 
    The relatively weak dispersion of $\alpha$ for the most massive ones on the other hand suggests that 
    %It has recently been argued \citep{ibanez2016} that the turbulence in molecular clouds should be attributed 
    %to gravity. Indeed the $\alpha$ values shown in \citet{ibanez2016} are notably below the ones obtained here. The reason is probably 
    %that in this study the only source of feedback are the supernovae, which are uncorrelated with the dense gas and therefore gravity 
    %is the only source of turbulence within dense clouds. Note that the galactic disc in these simulations 
    %is very narrow and the SFR presumably  very high. 
    they are primarily driven by a combination of self-gravity and feedback. It is likely that
    for these objects, $\alpha$ tends to be self-regulated. 
    First of all, we may expect a selection effect. 
      While the most massive and unstable clouds are born out of an ensemble of structures,
    that on average are dominated by turbulence individually (as shown in Fig.~\ref{fig:alpha-mu}), 
    they would not have become strongly self-gravitating if $\alpha$ was too large. Second, there 
    is likely an evolutionary effect induced by gravity which tends to produce collapse motions, that 
    also have $\alpha$ on the order of a few although this may preferentially occur at scales 
    not well described in the present simulations. Finally, when feedback becomes strong enough to unbind the 
    clouds, that is to say when the diverging motions have $\alpha$ again equal to a few, the cloud is 
    disrupted.  These points are illustrated by Fig.~\ref{clump} that displays the 
    column density and the integrated velocity for four most massive clouds in simulations B1. 
    Only the most massive one (top left panel) shows some clear sign of global infall and even there, it is obvious that 
    there are plenty of non-infalling and disordered motions. For the three other clouds shown the most obvious 
    trends are the diverging motions which are due to supernova feedback. This implies that in the present simulations, 
    feedback processes, which are both spatially and temporally correlated with star formation, 
    start destroying the clouds before a global collapse takes place. Although this behaviour likely depends on 
    the details of the feedback processes which are not sufficiently accurately described in this work, it must be reminded that 
    the star formation efficiency is observed to be rather low in molecular clouds \citep{lada2010}, which is barely 
    compatible with global infall being dominant on large scales. At smaller scales, not well described in this work, the situation may be different
    \citep{peretto2007,ballesteros2011}.

    The mass-to-flux over critical mass-to-flux ratio, $\mu$, is also displayed in Fig.~\ref{fig:alpha-mu}. It increases 
    from values of about 0.3 for 100 M$_\odot$ clouds to about 8-10 for cloud masses of $10^4$ M$_\odot$ and
    presents a rough scaling $\mu \propto M^{1/2}$. A similar relation has been obtained by \citet{Banerjee09} and \citet{Inoue2012} 
    where a relation $\mu \propto M^{0.4}$  has been inferred. As magnetic field is playing a significant role, it is important to 
    understand the origin of this relation keeping in mind that getting the normalization factor (that is to say the value of 
    $\mu$ at some specific value) is not straightforward,
    because, as discussed above, magnetic flux is getting expelled from the galactic disc. 
    Let us consider that the gas has reached some statistical equilibrium  constituted by a mixture of 
    warm and cold dense gas, that it has a mean density $\rho_{ISM}$ and that it is threaded by a mean magnetic field,  $B_{ISM}$. 
    As a structure gets assembled out of a radius $R$, its mass, $M$, is typically  $\propto \rho_{ISM} R^3$, 
    while the flux is $\propto R^2 B _{ISM}$. Therefore the mass-to-flux ratio, $\mu$, is thus $\propto R$. Since we get a 
    mass size relation $M \propto R^{2.3}$, we get 
    \begin{eqnarray}
      \mu \propto M/\Phi \propto R \propto M^{1/2.3} \simeq M^{0.43},
    \end{eqnarray}
    which is in good agreement with the observed 
    behaviour. This relation is of importance as it leads to a prediction  of the field intensity in ISM structures. 
    As we see it is essentially due to a simple geometrical effect, larger structures having a larger volume over surface ratio
    than smaller ones.

    Finally Fig.~\ref{fig:mclump} shows the mass spectrum of the clumps for all simulations. 
    The shape observed in smaller scale simulations is recovered 
    \citep[e.g.][]{Hennebelle2007,Heitsch2008,Banerjee09,Inoue2012,Padoan16,Valdivia16}
    with a  plateau at small masses and a power-law at high masses. While the 
    latter is a consequence of numerical dissipation, the former likely reflects the 
    properties of turbulence as discussed in \citet{HC2008}.
    This good agreement between simulations performed at scales of 50 pc and the present ones
    which resolve the galactic disc is in good agreement with the idea that a large scale turbulent 
    cascade is taking place and that the limited range of structure distribution, a clear 
    consequence, of limited resolution, can be extrapolated to the regime of smaller structures. 

    Figure~\ref{fig:mclump} confirms that stronger fields tend to diminish the number of small scale 
    structures (see for example runs B0W, B1W and B2W which all have a resolution of 1024$^3$).
    Interestingly we also see that  the power-law becomes  shallower when the magnetic intensity 
    increases going from about $dN/dM \propto M^{-2}$ in the hydrodynamical simulations (run B0) to 
     $dN/dM \propto M^{-1.5}$ (run B4). This is also consistent with the idea that the fluid particle  being partially 
    linked by the field lines, they tend to form bigger clumps. Observationally 
    a slope of about 1.7 has been inferred from CO survey \citep[e.g.][]{Kramer98,Heithausen98}.
    Since run B1 presents an exponent close to this value, this is consistent 
     as the large scale  magnetic field in this run is on the order of 3 $\mu G$ and is 
     therefore close to the mean galactic field.

%__________________________________________________________________

\section{Conclusion}

We have performed a series of high resolution tridimensional numerical simulations 
with a resolution up to 1024$^3$, aiming 
to describe self-consistently the vertical structure of a galactic disc and a
self-regulated star-forming ISM through supernova feedback. 
We considered four magnetizations and two feedback injections, one 
using canonical  momentum injected by the supernovae and one 
four  times below this value. 

The measured SFR are comparable to the observational values, particularly 
with the standard feedback and magnetization. It is roughly four times larger 
when the weak feedback scheme is used. The hydrodynamical runs present
SFR two times larger than the intermediate magnetization and the 
run with the strongest field presents SFR 2-3 times lower than in the 
intermediate field case. We found that while significant, the impact of the magnetic field 
tends to be limited by two effects. First of all magnetic flux tends to be expelled 
from the galactic plane probably because of the turbulent motions arising there. Second 
of all the magnetic and velocity fields are preferentially aligned reducing the 
effect of the Lorentz force. 
Comparison between an analytical model and the measured scale height, shows that indeed, 
except for the most magnetized runs, the magnetic field does not increase the 
disc scale height significantly. This allows us to also estimate the 
efficiency of the energy injection by the supernovae onto the gas within the galactic disc 
and we find it to be on the order of a few percents. 

We computed  tridimensional power spectra of various flow quantities such as  density, 
magnetic field and velocity, finding classical behaviour although the slopes 
are closer to the canonical 11/3 than the 3.9 inferred for supersonic turbulence. 
As the simulations are strongly stratified, we also computed 
bidimensional power spectra in a series of horizontal planes at various heights. In 
particular, we performed a Helmholtz decomposition and found that in the 
equatorial plane, even for the strongly magnetized runs, 
the compressible modes tend to dominate the solenoidal ones. 
At higher heights the former becomes negligible. 
We stress that the dominance of the compressible modes in the galactic plane is possibly 
biased by our particular choice of supernovae driving.

Finally, we extracted the dense clouds and computed their physical properties, finding 
them to be reminiscent of the observed clouds though we do not exclude that their 
internal velocities may be too low, which may indicate that either feedback is not 
strong enough, either there is further energy injection from the large 
galactic scales. The mass-to-flux ratio is found to be $\propto M^{0.4-0.5}$
and a simple explanation has been proposed. The virial parameter, $\alpha$, 
has been estimated and the shape of mass$-\alpha$ distribution is also similar 
to observations. At masses $M \simeq 10^{2-3}$ M$_\odot$ $\alpha$ presents a large spread and is typically 
equal to  10-100.  At masses $M \simeq 10^{4-5}$ M$_\odot$ $\alpha$ is of the order of a few 
and presents a narrow distribution.

%__________________________________________________________________

\begin{acknowledgements}
%\emph{Acknowledgments}
We thank the anonymous referee for a thorough and constructive report, which has 
improved the manuscript.
This work was granted access to HPC resources of CINES and IDRIS  under the 
allocation  x2016047023 made by GENCI (Grand Equipement National de Calcul Intensif).
PH acknowledge the finantial support of the Agence Nationale pour la Recherche through the 
 COSMIS project.
This research has received funding from the European Research Council under the European
 Community's Seventh Framework Programme (FP7/2007-2013 Grant Agreement no. 306483).

\end{acknowledgements}

% for the bibliography
\bibliography{refs}{}
\bibliographystyle{aa} % style aa.bst

\begin{appendix}

\section{Three-dimensional power spectra of high resolution runs}
\label{appen-3D}

         \begin{figure}
            \begin{center}
               \includegraphics[width=8cm]{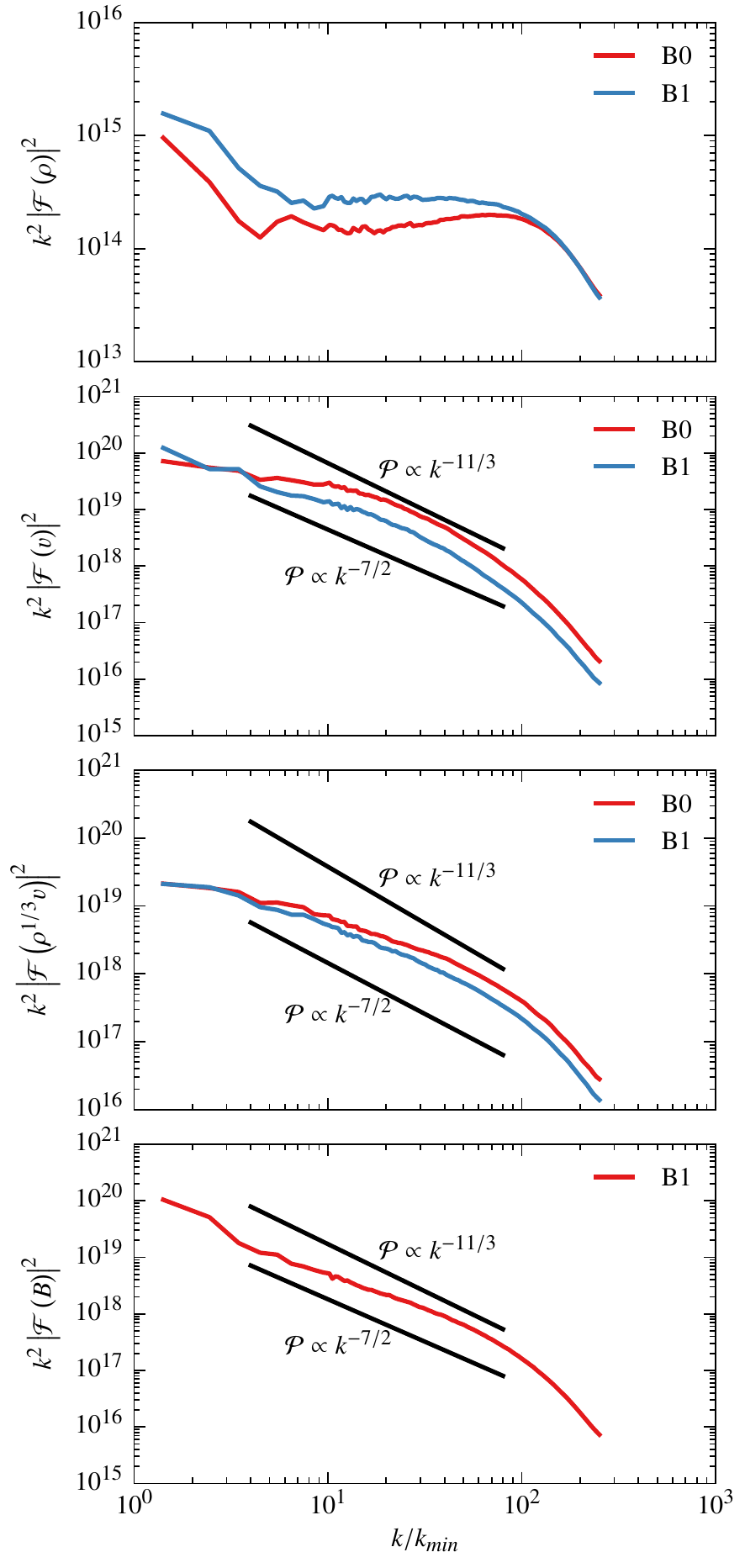}
            \end{center}
         \caption{Runs B0 and B1 (high resolution with standard feedback). Three-dimensional power spectra.
         \emph{From top to bottom:} density, velocity, density-weighted
         velocity, magnetic field. %and log of density.
         The spectra are multiplied by $k^2$ (such that the Kolmogorov scaling
         corresponds to a slope of $-5/3$)}%
         \label{fig:pspec_h_3d_s}
         \end{figure}

         \begin{figure}
            \begin{center}
               \includegraphics[width=8cm]{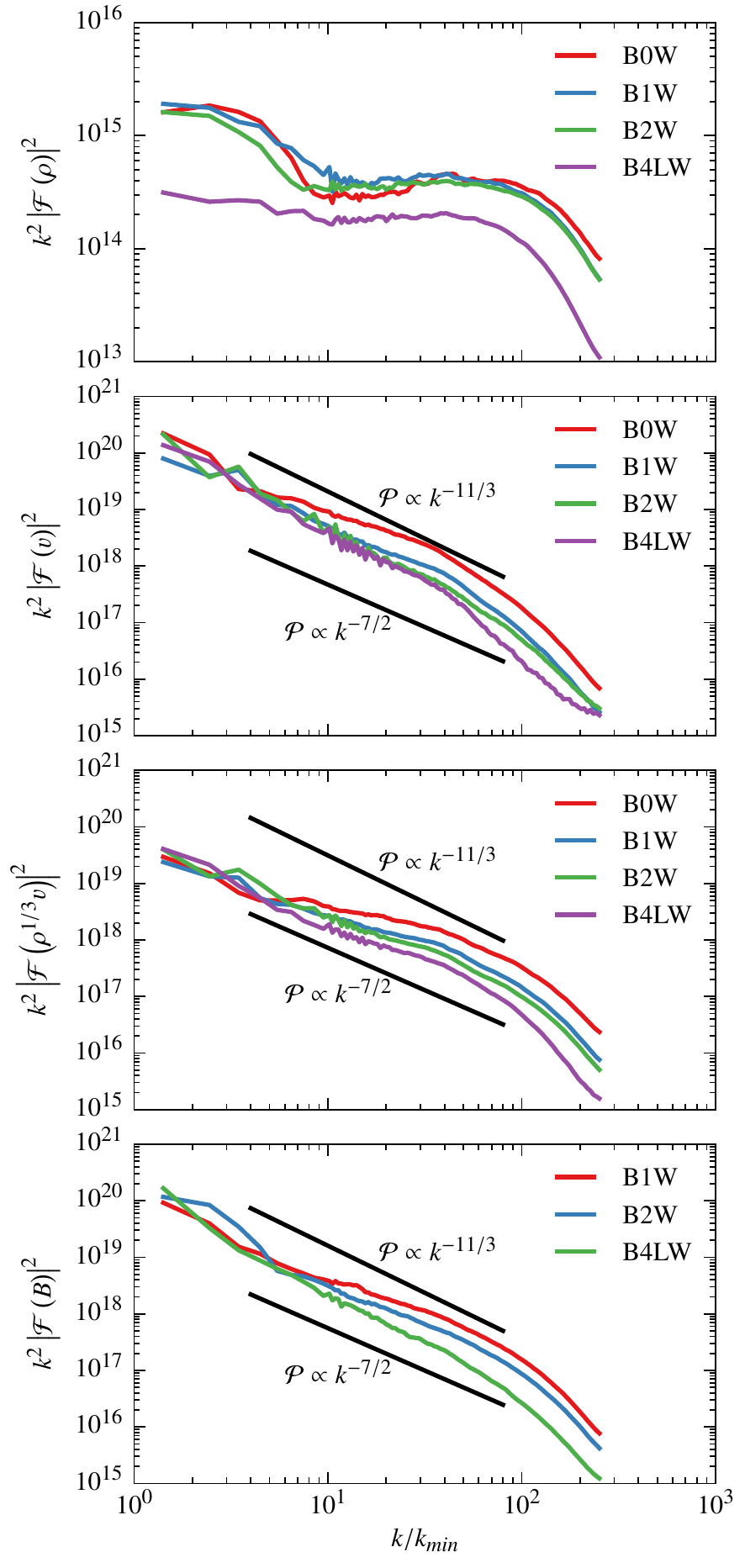}
            \end{center}
         \caption{Runs with weak feedback (B0W, B1W, B2W and B4LW). Three-dimensional power spectra.
         \emph{From top to bottom:} density, velocity, density-weighted
         velocity, magnetic field. % and log of density.
         The spectra are multiplied by $k^2$ (such that the Kolmogorov scaling
         corresponds to a slope of $-5/3$)}%
         \label{fig:pspec_w_3d_s}
         \end{figure}

Here we provide for comparison and reference the power spectra of the high resolution runs
corresponding to the standard feedback (Fig.~\ref{fig:pspec_h_3d_s}) 
and to the weak feedback (Fig.~\ref{fig:pspec_w_3d_s}). 
As can be seen these spectra are very similar to the 
ones presented in Fig.~\ref{fig:pspec_m_3d_s}, which shows that reasonable convergence has been reached
but also that the weak feedback does not affect too much the flow properties.

\section{Two-dimensional power spectra of density and magnetic field}
\label{appen-2D}

         \begin{figure*}
            \begin{center}
              \includegraphics[width=17cm]{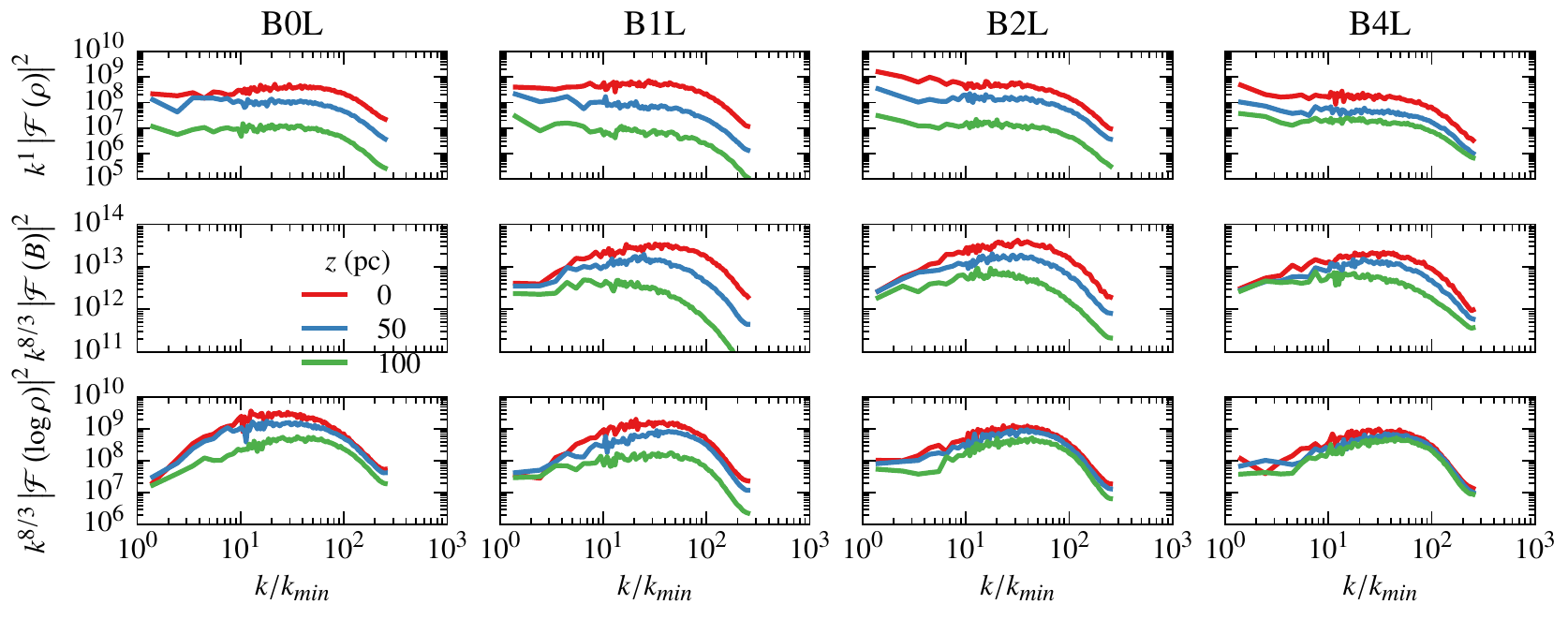}
            \end{center}
         \caption{Two-dimensional velocity power spectra of $\rho$, $B$ and $\log \rho$.
         \emph{From left to right:} runs B0L, B1L, B2L, and B4L.
         \emph{From top to bottom:} altitude $0$, $50$, and $100\ \mathrm{pc}$.
         The spectra are multiplied by $k^{8/3}$ for comparison with the Kolmogorov
         scaling law.}%
         \label{fig:pspec_mult_2d_s}
         \end{figure*}

As the simulations presented here have a strong stratification, we show for the 
sake of completeness a series of two-dimensional power spectra obtained 
at three altitudes. As can be seen the index of the power spectra are broadly compatible
to the three-dimensional ones presented  in Fig.~\ref{fig:pspec_m_3d_s} with some 
noticeable difference. In particular the index of the magnetic field power spectra 
varies with altitude.

\section{Alignment of velocity and magnetic fields for  density range}
\label{BV_rho}

   \begin{figure*}
      \begin{center}
         \includegraphics[width=17cm]{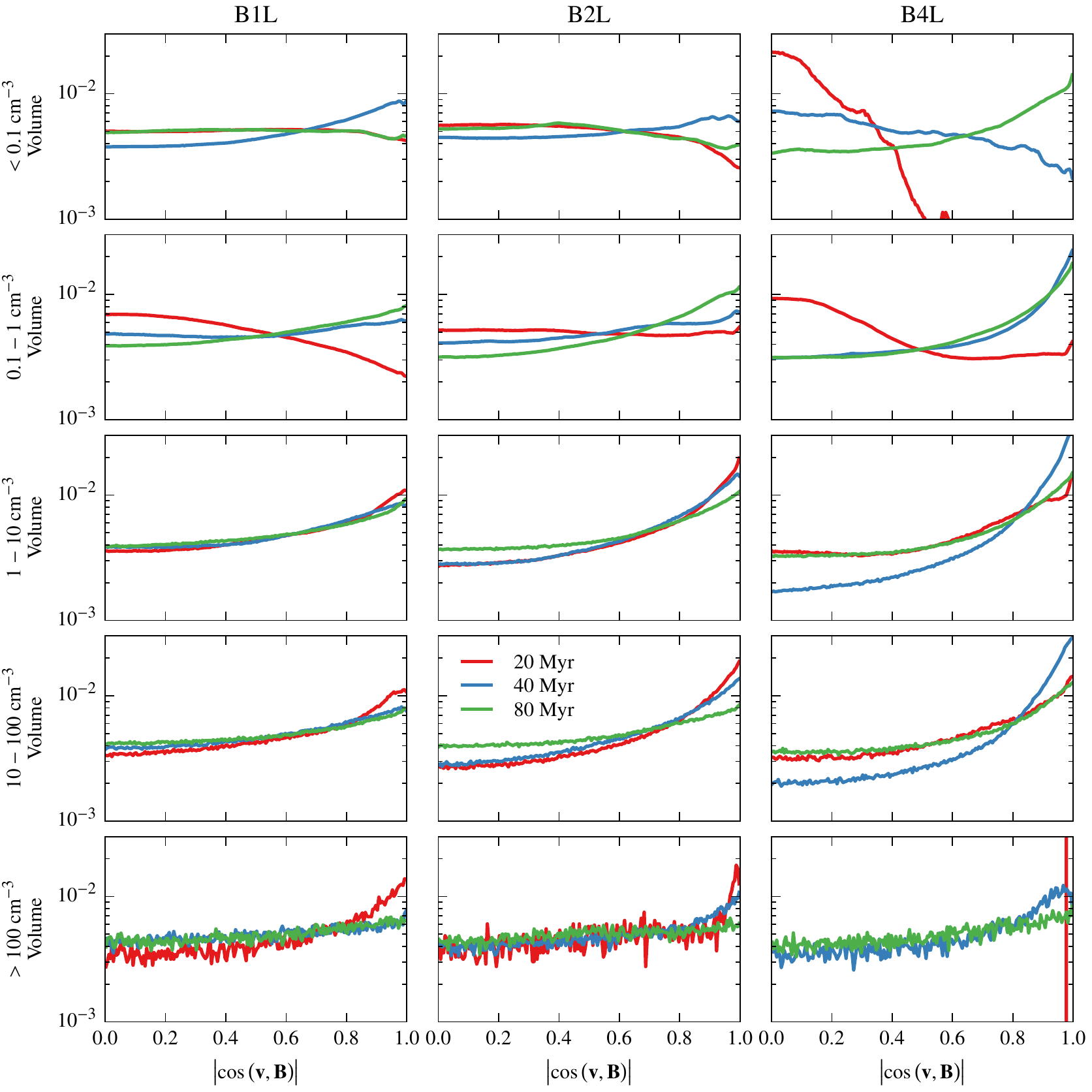} \\[1ex]
      \end{center}
   \caption{Relative orientation of the velocity and the magnetic fields
     as a function of time for five density bins. 
   }\label{fig:BV_rho}
   \end{figure*}

Figure~\ref{fig:BV_rho} displays the relative orientation of the velocity and 
magnetic fields as a function of time and for five density bins allowing a 
more detailed analysis than Fig.~\ref{fig:v_B_angle_w}.  
As can be seen in the very diffuse gas ($n<0.1$ cm$^{-3}$), there is 
no alignment. Clearly this is because this gas is produced by supernova 
explosions. For denser gas, the relative orientation distribution is 
nearly the same for the three bins 0.1-1, 1-10 and 10-100 cm$^{-3}$. 
As expected the alignment is stronger when the field intensity is higher. 
Interestingly, there is a clear trend for the alignment being less 
pronounced for $n > 100$ cm$^{-3}$.
      This is consistent with the contraction occurring mainly along field lines 
      at low and intermediate densities and becoming less focused at higher densities, 
      either because gravity leads to global contraction or because the Alfv\'enic 
      Mach numbers tends to be higher at higher densities.

\section{Velocity as a function of $R \Sigma$}
\label{appen-sig-Rsig}

   \begin{figure*}
      \begin{center}
         \includegraphics[width=17cm]{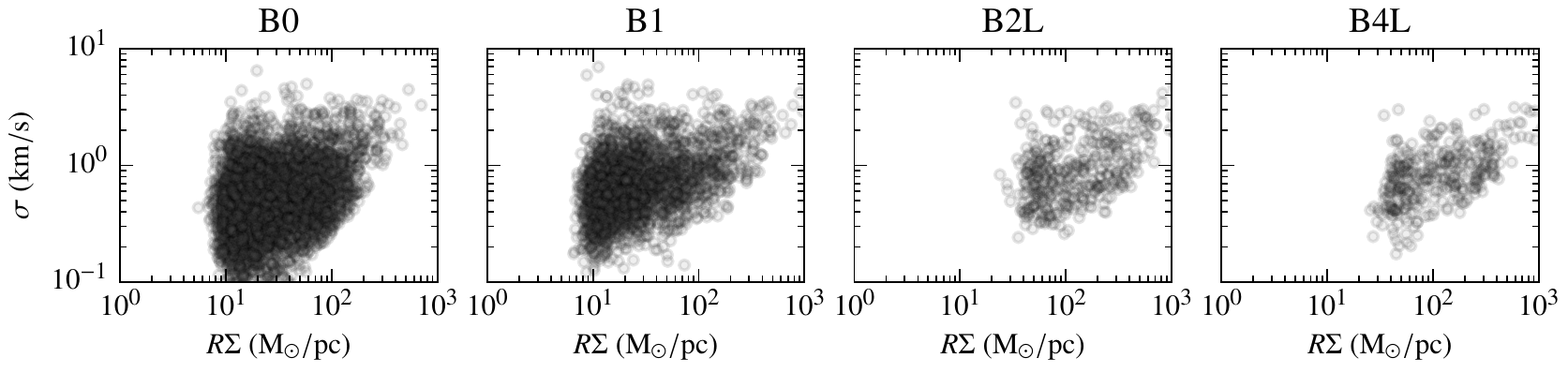} \\[1ex]
         \includegraphics[width=17cm]{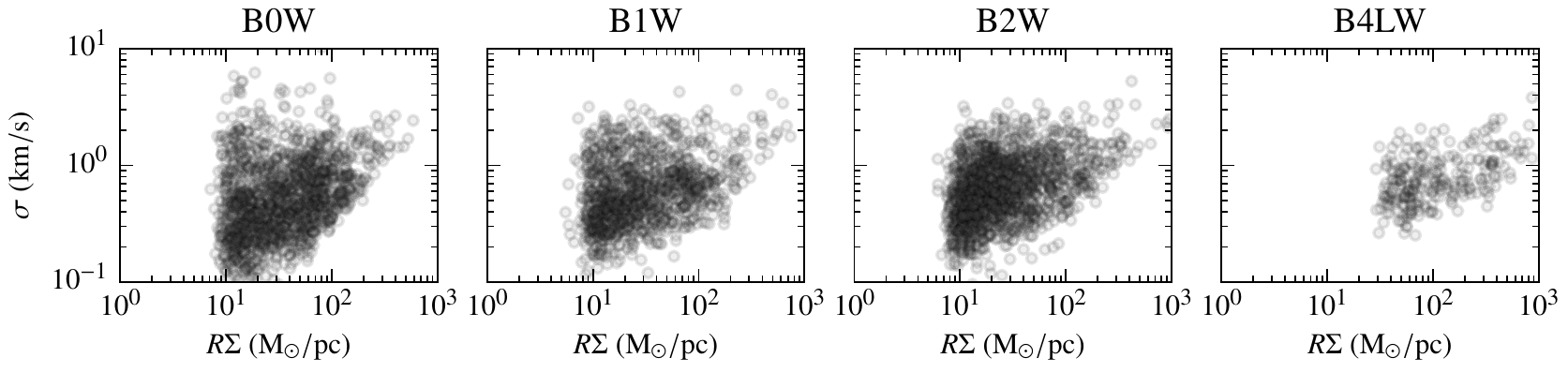}
      \end{center}
   \caption{Clump scaling relations at $60\ \mathrm{Myr}$.
   \emph{From left to right:} runs B0, B1, B2, and B4.
   size times column density-velocity  dispersion relation.
   \emph{Top panels:} strong feedback. \emph{Bottom panels:} weak feedback.
   }\label{fig:sig-RSig}
   \end{figure*}

Figure~\ref{fig:sig-RSig} shows the velocity dispersion as a function of 
$R \Sigma$ for comparisons with \citet{miville2016}. The numbers inferred are 
pretty similar. However, the correlation is not obviously better than 
the $\sigma-R$ one displayed in Fig.~\ref{fig:larson}.

\end{appendix}

\end{document}